# Actor Model of Computation:
# Scalable Robust Information Systems

**Carl Hewitt**

*This article is dedicated to Alonzo Church and Dana Scott.*

The Actor Model is a mathematical theory that treats "*Actors*" as the universal primitives of digital computation.

Hypothesis:[i] **All physically possible computation can be directly implemented using Actors.**

The model has been used both as a framework for a theoretical understanding of concurrency, and as the theoretical basis for several practical implementations of concurrent systems. The advent of massive concurrency through client-cloud computing and many-core computer architectures has galvanized interest in the Actor Model.

**Message passing using types is the foundation of system communication:**
- Messages are the unit of communication[1]
- Types enable secure communication with any Actor

When an Actor receives a message, it can concurrently:
- send messages to (unforgeable) addresses of Actors that it has;
- create new Actors;
- designate how to handle the next message it receives.

The Actor Model can be used as a framework for modeling, understanding, and reasoning about, a wide range of concurrent systems. For example:
- Electronic mail (e-mail) can be modeled as an Actor system. Mail accounts are modeled as Actors and email addresses as Actor addresses.
- Web Services can be modeled with endpoints modeled as Actor addresses.
- Objects with locks (e.g. as in Java and C#) can be modeled as Actors.
- Functional and Logic programs can be implemented using Actors.

Actor technology will see significant application for integrating all kinds of digital information for individuals, groups, and organizations so their information usefully links together.

---

[i] This hypothesis is an update to [Church 1936] that all physically computable functions can be implemented using the lambda calculus. It is a consequence of the Actor Model that there are some computations that *cannot* be implemented in the lambda calculus.

Information integration needs to make use of the following information system principles:

- **Persistence**. *Information is collected and indexed.*
- **Concurrency**: *Work proceeds interactively and concurrently, overlapping in time.*
- **Quasi-commutativity**: *Information can be used regardless of whether it initiates new work or become relevant to ongoing work.*
- **Sponsorship**: *Sponsors provide resources for computation, i.e., processing, storage, and communications.*
- **Pluralism**: *Information is heterogeneous, overlapping and often inconsistent. There is no central arbiter of truth.*
- **Provenance**: *The provenance of information is carefully tracked and recorded.*

The Actor Model is intended to provide a foundation for inconsistency robust information integration. Inconsistency[i] robustness is information system performance[2] in the face of continual pervasive inconsistencies---a shift from the previously dominant paradigms of inconsistency denial[3] and inconsistency elimination attempting to sweep inconsistencies under the rug. Inconsistency robustness is both an observed phenomenon and a desired feature.

The Actor Model is a mathematical theory of computation that treats "*Actors*" as the universal primitives of concurrent digital computation [Hewitt, Bishop, and Steiger 1973; Hewitt 1977]. The model has been used both as a framework for a theoretical understanding of concurrency, and as the theoretical basis for several practical implementations of concurrent systems.

Unlike previous models of computation, the Actor Model was inspired by physical laws. It was also influenced by the programming languages Lisp [McCarthy *et. al.* 1962], Simula-67 [Dahl and Nygaard 1967] and Smalltalk-72 [Kay 1975], as well as ideas for Petri Nets [Petri 1962], capability systems [Dennis and van Horn 1966] and packet switching [Baran 1964]. The advent of massive concurrency through client-cloud computing and many-core computer architectures has galvanized interest in the Actor Model [Hewitt 2009b].

---

[i] An inference system is *inconsistent* when it is possible to derive both a proposition and its negation.

A *contradiction* is manifest when both a proposition and its negation are asserted even if by different parties, *e.g.*, New York Times said "*Snowden is a whistleblower.*", but NSA said "*Snowden is not a whistleblower.*"

It is important to distinguish the following:
- modeling arbitrary computational systems using Actors.[i] It is difficult to find physical computational systems (regardless of how idiosyncratic) that cannot be modeled using Actors.
- securely implementing practical computational applications using Actors remains an active area of research and development.

## Fundamental concepts

An Actor receives a messages, it can concurrently:[4]
- send messages to (unforgeable) addresses of Actors;
- create new Actors[ii]
- designate how to handle the next message it receives.

Decoupling the sender from the communications it sends was a fundamental advance of the Actor Model enabling asynchronous communication and control structures as patterns of passing messages [Hewitt 1977].

An Actor can only communicate with another Actor to which it has an address.[iii] Addresses can be implemented in a variety of ways:
- direct physical attachment
- memory or disk addresses
- network addresses
- email addresses

The Actor Model is characterized by inherent concurrency of computation within and among Actors, dynamic creation of Actors, inclusion of Actor addresses in messages, and interaction only through direct asynchronous message passing with no restriction on message reception order.

The Actor Model differs from its predecessors and most current models of computation in that the Actor Model assumes the following:
- Concurrent execution in processing a message.
- The following are *not* required by an Actor: a thread, a mailbox, a message queue, its own operating system process, *etc.*[iv]

---

[i] An Actor can be implemented directly in hardware.

[ii] with new addresses

[iii] In the literature, an Actor address is sometimes called a "capability"[Dennis and van Horn 1966] because it provides the capability to send a message.

[iv] For example, if an Actor were required to have a mailbox then, the mailbox would be an Actor that is required to have its own mailbox…

- Message passing has the same overhead as looping and procedure calling.
- Primitive Actors can be implemented in hardware.[i]

The Actor Model can be used as a framework for modeling, understanding, and reasoning about, a wide range of concurrent systems.

For example:
- Electronic mail (e-mail) can be modeled as an Actor system. Mail accounts are modeled as Actors and email addresses as Actor addresses.
- Web Services can be modeled with SOAP endpoints modeled as Actor addresses.
- Objects with locks (*e.g.* as in Java and C#) can be modeled as Actors.

## Direct communication and asynchrony

The Actor Model is based on one-way asynchronous communication. Once a message has been sent, it is the responsibility of the receiver.[5]

Messages in the Actor Model are decoupled from the sender and are delivered by the system on a best efforts basis.[6] This was a sharp break with previous approaches to models of concurrent computation in which message sending is tightly coupled with the sender and sending a message synchronously transfers it someplace, *e.g.*, to a buffer, queue, mailbox, channel, broker, server, *etc.* or to the "*ether*" or "*environment*" where it temporarily resides. The lack of synchronicity caused a great deal of misunderstanding at the time of the development of the Actor Model and is still a controversial issue.

Because message passing is taken as fundamental in the Actor Model, there cannot be any required overhead, *e.g.,* any requirement to use buffers, pipes, queues, classes, channels, *etc.* Prior to the Actor Model, concurrency was defined in low level machine terms.

It certainly is the case that implementations of the Actor Model typically make use of these hardware capabilities. However, there is no reason that the model could not be implemented directly in hardware without exposing any hardware threads, locks, queues, cores, channels, tasks, *etc.* Also, there is no necessary relationship between the number of Actors and the number threads, cores, locks, tasks, queues, *etc.* that might be in use. Implementations of the Actor Model are free to make use of threads, locks, tasks, queues, global, coherent memory, transactional memory, cores, *etc.* in any way that is compatible with the laws for Actors [Baker and Hewitt 1977].

---

[i] In some cases, this involves (clocked) one-way messages so message guarantees and exception processing can be different from typical application Actors.

As opposed to the previous approach based on composing sequential processes, the Actor Model was developed as an inherently concurrent model. In the Actor Model sequential ordering is a special case that derived from concurrent computation. Also, the Actor Model is based on communication rather that a global state with an associated memory model as in Turing Machines, CSP [Hoare 1978], Java [Sun 1995, 2004], C++11 [ISO 2011], X86 [AMD 2011], *etc.*

A natural development of the Actor Model was to allow Actor addresses in messages. A computation might need to send a message to a recipient from which it would later receive a response. The way to do this is to send a communication which has the message along with the address of another Actor called the *customer* along with the message. The recipient could then cause a response message to be sent to the customer.

Of course, any Actor could be used as a customer to receive a response message. By using customers, common control structures such a recursion, co-routines, hierarchical parallelism, futures [Baker and Hewitt 1977, Hewitt 2011], *etc.* can be implemented.

### Indeterminacy and Quasi-commutativity

The Actor Model supports indeterminacy because the reception order of messages can affect future behavior.

Operations are said to be quasi-commutative to the extent that it doesn't matter in which order they occur. To the extent possible, quasi-commutativity is used to reduce indeterminacy.

### Locality and Security

Locality and security are important characteristics of the Actor Model[Baker and Hewitt 1977].[7]

Locality and security mean that in processing a message: an Actor can send messages only to addresses for which it has information by the following means:

1. that it receives in the message
2. that it already had before it received the message
3. that it creates while processing the message.

In the Actor Model, there is no hypothesis of simultaneous change in multiple locations. In this way it differs from some other models of concurrency, *e.g.*, the Petri net model in which tokens are simultaneously removed from multiple locations and placed in other locations.

The security of Actors can be protected in the following ways:
- hardwiring in which Actors are physically connected
- every-word-tagged memory.
- virtual machines as in Java virtual machine, Common Language Runtime, *etc.*
- signing and/or encryption of Actors and their addresses

A delicate point in the Actor Model is the ability to synthesize the address of an Actor. In some cases security can be used to prevent the synthesis of addresses in practice using the following:
- every-word-tagged memory
- signing and encryption of messages

## Robustness in Runtime Failures

Runtime failures are always a possibility in Actor systems and are dealt with by runtime infrastructures. Message acknowledgement, reception, and response[i] cannot be guaranteed although best efforts are made. Consequences are cleaned up on a best-effort basis.

Robustness is based on the following principle:

If an Actor is sent a request, then the continuation *will* be one of the following two mutually exclusive possibilities:
1. to respond with the response received from the Actor sent the request
2. to throw a Messaging[ii] exception[iii]

## Scalability and Modularity

ActorScript™ is a general purpose programming language for implementing iAdaptive™ concurrency that manages resources and demand. It is differentiated from previous languages by the following:
- Universality
  - Ability to directly specify what Actors can do
  - Specify interface between hardware and software
  - Everything in the language is accomplished using message passing including the very definition of ActorScript itself.

---

[i] a response is either a returned value or a thrown exception

[ii] A Messaging exception can have information concerning the lack of response

[iii] even though the Actor may have received the request and sent a response that has not yet been received. Requestors need to be able to interact with infrastructures concerning policies to be applied concerning when to generate Unresponsive exceptions.

- o Functional, Imperative, Logic, and Concurrent programming are integrated. Concurrency can be dynamically adapted to resources available and current load.
- o Programs do not expose low-level implementation mechanisms such as threads, tasks, channels, coherent memory, location transparency, throttling, load balancing, locks, cores, *etc*. Messages can be directly communicated without requiring indirection through brokers, channels, class hierarchies, mailboxes, pipes, ports, queues *etc*. Variable races are eliminated.
- o Binary XML and JSON are data types.
- o Application binary interfaces are afforded so that no program symbol need be looked up at runtime.

- Safety and security
  - o Programs are extension invariant, i.e., extending a program does not change its meaning.
  - o Applications cannot directly harm each other.
- Performance
  - o Impose no overhead on implementation of Actor systems
  - o Message passing has essentially same overhead as procedure calling and looping.
  - o Execution dynamically adjusted for system load and capacity (*e.g.* cores)
  - o Locality because execution is not bound by a sequential global memory model
  - o Inherent concurrency because execution is not bound by communicating sequential processes
  - o Minimize latency along critical paths

ActorScript attempts to achieve the highest level of performance, scalability, and expressibility with a minimum of primitives.

## Scalable information integration

Technology now at hand can integrate all kinds of digital information for individuals, groups, and organizations so their information usefully links together. This integration can include calendars and to-do lists, communications (including email, SMS, Twitter, Facebook), presence information (including who else is in the neighborhood), physical (including GPS recordings), psychological (including facial expression, heart rate, voice stress) and social (including family, friends, team mates, and colleagues), maps (including firms, points of interest, traffic, parking, and weather), events (including alerts and status), documents (including presentations, spreadsheets, proposals, job applications, health records, photos, videos, gift lists, memos, purchasing, contracts, articles), contacts (including social graphs and reputation), purchasing information (including store purchases, web purchases, GPS and phone records, and buying and travel habits), government information (including licenses, taxes, and rulings), and search results (including rankings and ratings).

## Connections

Information integration works by making connections including examples like the following:

- A statistical connection between "being in a traffic jam" and "driving in downtown Trenton between 5PM and 6PM on a weekday."
- A terminological connection between "MSR" and "Microsoft Research."
- A causal connection between "joining a group" and "being a member of the group."
- A syntactic connection between "a pin dropped" and "a dropped pin."
- A biological connection between "a dolphin" and "a mammal".
- A demographic connection between "undocumented residents of California" and "7% of the population of California."
- A geographical connection between "Leeds" and "England."
- A temporal connection between "turning on a computer" and "joining an on-line discussion."

By making these connections iInfo[TM] information integration offers tremendous value for individuals, families, groups, and organizations in making more effective use of information technology.

## Information Integration Principles

In practice integrated information is invariably inconsistent.[8] Therefore iInfo must be able to make connections even in the face of inconsistency.[9] The business of iInfo is not to make difficult decisions like deciding the ultimate truth or probability of propositions. Instead it provides means for processing information and carefully recording its provenance including arguments (including arguments about arguments) for and against propositions.

Information integration needs to make use of the following principles:

- **Persistence**. *Information is collected and indexed and no original information is lost.*
- **Concurrency**: *Work proceeds interactively and concurrently, overlapping in time.*
- **Quasi-commutativity**: *Information can be used regardless of whether it initiates new work or become relevant to ongoing work.*
- **Sponsorship**: *Sponsors provide resources for computation, i.e., processing, storage, and communications.*
- **Pluralism**: *Information is heterogeneous, overlapping and often inconsistent. There is no central arbiter of truth*
- **Provenance**: *The provenance of information is carefully tracked and recorded*

**Interaction creates Reality**[10]

> *a philosophical shift in which knowledge is no longer treated primarily as referential, as a set of statements **about** reality, but as a practice that interferes with other practices. It therefore participates **in** reality.*
>   Annemarie Mol [2002]

Relational physics takes the following view [Laudisa and Rovelli 2008]:[i]
- Relational physics discards the notions of absolute state of a system and absolute properties and values of its physical quantities.
- State and physical quantities refer always to the interaction, or the relation, among multiple systems.
- Nevertheless, relational physics is a complete description of reality.

According to this view, **Interaction creates reality.** Information systems participate in this reality and thus are both consequence and cause.

Actor systems can be organized in higher level structures to facilitate operations.

## Organizational Programming using iOrgs

The Actor Model supports Organizational Programming that is based on authority and accountability in iOrgs [Hewitt 2008a] with the goal of becoming an effective readily understood approach for addressing scalability issues in Software Engineering. The paradigm takes its inspiration from human organizations. iOrgs provide a framework for addressing issues of hierarchy, authority, accountability, scalability, and robustness using methods that are analogous to human organizations. Because humans are very familiar with the principles, methods, and practices of human organizations, they can transfer this knowledge and experience to iOrgs. iOrgs achieve scalability using methods and principles similar to those used in human organizations. For example an iOrg can have sub-organizations specialized by areas such as sales, production, and so forth. Authority is delegated down the organizational structure and when necessary issues are escalated upward. Authority requires accountability for its use including record keeping and periodic reports. Management is in large part the art of reconciling authority and accountability.

---

[i] According to [Rovelli 1996]:  *Quantum mechanics is a theory about the physical description of physical systems relative to other systems, and this is a complete description of the world.*

[Feynman 1965] offered the following advice:  *Do not keep saying to yourself, if you can possibly avoid it, "But how can it be like that?" because you will go "down the drain," into a blind alley from which nobody has yet escaped.*

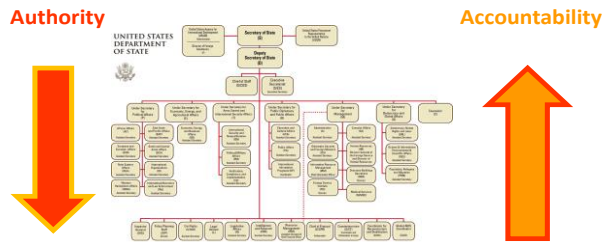

## Organizational Programming for iOrgs

iOrgs are structured around *organizational commitment* defined as information pledged constituting an alliance to go forward. For example, iOrgs can use contracts to formalize their mutual commitments to fulfill specified obligations to each other.

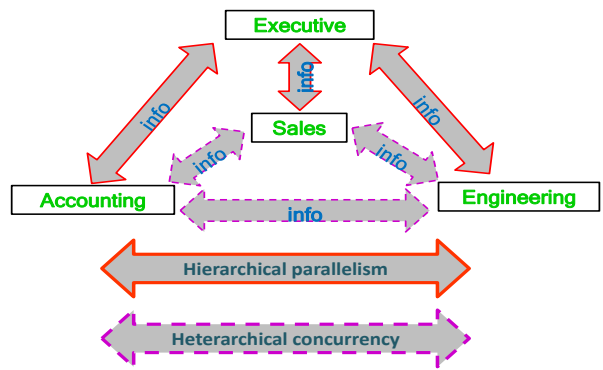

## Scalability of iOrgs

Yet, manifestations of information pledged will often be inconsistent. Any given agreement might be internally inconsistent, or two agreements in force at one time could contradict each other.

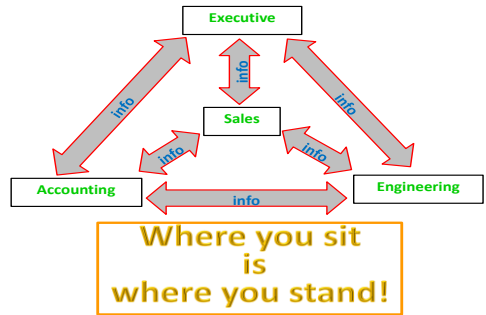

**Inconsistency by Design for iOrgs**

Issues that arise from such inconsistencies can be negotiated among iOrgs. For example the Sales department might have a different view than the Accounting department as to when a transaction should be booked.

A fundamental goal of Inconsistency Robustness is to effectively reason about large amounts of information at high degrees of abstraction:

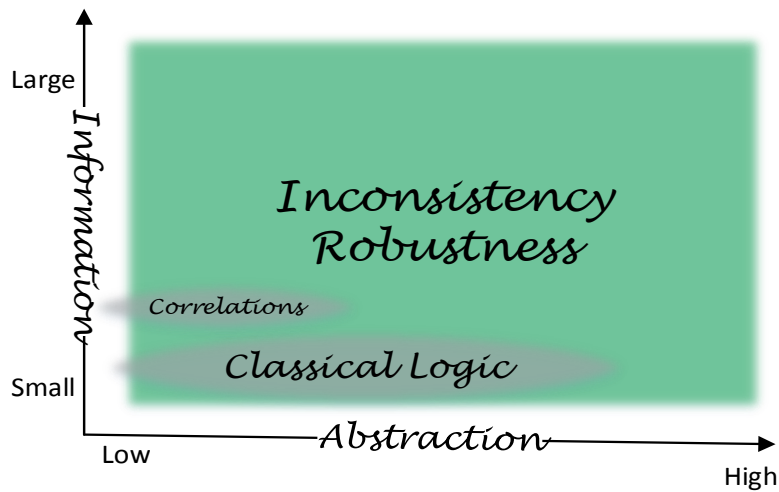

## Actor Implementations

Below is a simple implementation of Account:

Actor Account‹aCurrency⊑Currency›[startingBalance:aCurrency]
  // aCurrency must be a subtype of Currency, which is non-negative
  currentBalance := startingBalance
  getBalance[ ]:aCurrency → currentBalance
  deposit[anAmount:aCurrency]:Void →
    Void afterward currentBalance := currentBalance+anAmount
      // currentBalance is updated for the *next* message received
  withdraw[anAmount:aCurrency]:Void →
    anAmount>currentBalance ⍟
      True ⦂ Throw OverdrawnException[ ]
      False ⦂ Void afterward currentBalance := currentBalance-anAmount

The operations of reading the balance and making withdrawals are *quasi-commutative*[i]. For example the following expression returns €3 even though the withdrawals can occur in either order:

Let anAccount ← Account‹Euro›[€6]
  Do ⑩anAccount.withdraw[€1],        // concurrently withdraw €1 and €2
    ⑩anAccount.withdraw[€2];
                    // proceed only after both withdrawals have completed
  anAccount.getBalance[ ]

## Internet of Things

The Actor Model can help with the standardization of the Internet of Things (IoT). For example the above implementation can have the following interface description:

```
<Interface name="Account">
  <parameters>
     <type subtypeOf="Currency"> aCurrency</type>
  </parameters>
  <handler name="getBalance">
    <arguments/>
    <returns>aCurrency</returns>
  </handler>
  <handler name="withdraw">
    <arguments>aCurrency</arguments>
    <returns>Void</returns>
  </handler>
  <handler name="deposit">
    <arguments>aCurrency</arguments>
    <returns>Void</returns>
  </handler>
</Interface>
```

---

[i] Operations are quasi-commutative to the extent that it doesn't matter in which order they occur. Quasi-commutativity can be used to tame indeterminacy.

**Computational Representation Theorem**

The *Computational Representation Theorem* [Clinger 1981; Hewitt 2006][11] characterizes computation for systems which are closed in the sense that they do not receive communications from outside:

> The denotation $\mathbf{Denote_S}$ of a closed system $\mathbf{S}$ represents all the possible behaviors of $\mathbf{S}$ as
>
> $$\mathbf{Denote_S} \ = \ \lim_{i \to \infty} \text{Progressions}_S^{\ i}$$
>
> *where* $\text{Progressions}_S$ takes a set of partial behaviors to their next stage, i.e., $\text{Progressions}_S^i \blacktriangleright^i \text{Progressions}_S^{i+1}$

In this way, $\mathbf{S}$ can be mathematically characterized in terms of all its possible behaviors (including those involving unbounded nondeterminism).[ii]

> The denotations form the basis of constructively checking programs against all their possible executions,[iii]

A consequence of the Computational Representation Theorem is that there are uncountably many different Actors.

For example, $\text{Real}_{\bullet}[\ ]$ can output any real number between 0 and 1 where

$$\text{Real}_{\bullet}[\ ] \equiv [(0 \text{ either } 1), \mathbf{\forall}\text{Postpone } \text{Real}_{\bullet}[\ ]]$$

such that
- $(0 \text{ either } 1)$ is the nondeterministic choice of 0 or 1
- $[\text{first}, \mathbf{\forall}\text{rest}]$ is the list that begins with $\text{first}$ and whose remainder is $\text{rest}$
- $\text{Postpone}$ expression delays execution of expression until the value is needed.

The upshot is that ***concurrent systems can be axiomatized using mathematical logic[iv] but in general cannot be implemented***. Thus, the following practical problem arose:

> How can practical programming languages be rigorously defined since the proposal by Scott and Strachey [1971] to define them in terms lambda calculus failed because the lambda calculus cannot implement concurrency?[12]

A proposed answer to this question is the semantics of ActorScript [Hewitt 2010].

---

[i] read as "*can evolve to*"

[ii] There are no messages in transit in $\text{Denote}_S$

[iii] a restricted form of Model Checking in which the properties checked are limited to those that can be expressed in Linear-time Temporal Logic has been studied [Clarke, Emerson, Sifakis, *etc*. ACM 2007 Turing Award]

[iv] including the lambda calculus

### Extension versus Specialization

Programming languages like ActorScript [Hewitt 2010] take the approach of extending behavior in contrast to the approach of specializing behavior:

- Type specialization: If type t1 is a subtype of type t2, then instances of t1 have all of the properties that are provable from the definition of type t2 [Liskov 1987, Liskov and Wing 2001].
- Type extension: A type can be extended to have additional (perhaps incompatible) properties from the type that it extends. An extension type can make use of the implementation of the type that it extends. Type extension is commonly used to extend operating system software as well as applications.

The term "inheritance" in programming has been used (sometimes ambiguously) to mean both specialization and extension.

### Language constructs versus Library APIs

Library Application Programming Interfaces (APIs) are an alternative way to introduce concurrency.

For example,

- A limited version of futures[Baker and Hewitt 1977] have been introduced in C++11 [ISO 2011].
- Message Passing Interface (MPI) [Gropp et. al. 1998] provides some ability to pass messages.
- Grand Central Divide provides for queuing tasks.

There are a number of library APIs for Actor-like systems.

In general, appropriately defined language constructs provide greater power, flexibility, and performance than library APIs.[13]

### Reasoning about Actor Systems

The principle of Actor induction is:

1. Suppose that an Actor x has property P when it is created
2. Further suppose that if x has property P when it receives a message, then it has property P when it receives the next message.
3. Then x always has the property P.

In his doctoral dissertation, Aki Yonezawa developed further techniques for proving properties of Actor systems including those that make use of migration. Russ Atkinson developed techniques for proving properties of Actors that are guardians of shared resources. Gerry Barber's 1981 doctoral dissertation concerned reasoning about change in knowledgeable office systems.

**Other models of concurrency**

The Actor Model does not have the following restrictions of other models of concurrency:[14]

- *Single threadedness:* There are no restrictions on the use of threads in implementations.
- *Message delivery order:* There no restrictions on message delivery order.
- *Independence of sender:* The semantics of a message in the Actor Model is independent of the sender.
- *Lack of garbage collection (automated storage reclamation):* The Actor Model can be used in the following systems:
    - CLR and extensions (Microsoft and Xamarin)
    - JVM (Oracle and IBM)
    - LLVM (Apple)
    - Dalvik (Google)

    In due course, we will need to extend the above systems with a tagged extension of the X86 and ARM architectures. Many-core architecture has made a tagged extension necessary in order to provide the following:
    - concurrent, nonstop, no-pause automated storage reclamation (garbage collection) and relocation to improve performance,
    - prevention of memory corruption that otherwise results from programming languages like C and C++ using thousands of threads in a process,
    - nonstop migration of Actors (while they are in operation) within a computer and between distributed computers.

**Swiss Cheese**

Swiss cheese [Hewitt and Atkinson 1977, 1979; Atkinson 1980][15] is a programming language construct for scheduling concurrent access to shared resources with the following goals:

- *Generality:* Ability to conveniently program any scheduling policy
- *Performance:* Support maximum performance in implementation, *e.g.*, the ability to avoid repeatedly recalculating conditions for proceeding.
- *Understandability:* Invariants for the variables of an Actor should hold at all observable execution points.

Concurrency control for readers and writers in a shared resource is a classic problem. The fundamental constraint is that multiple writers are not allowed to operate concurrently and a writer is not allowed operate concurrently with a reader.

State diagram of **ReadersWriter** implementations:[i]

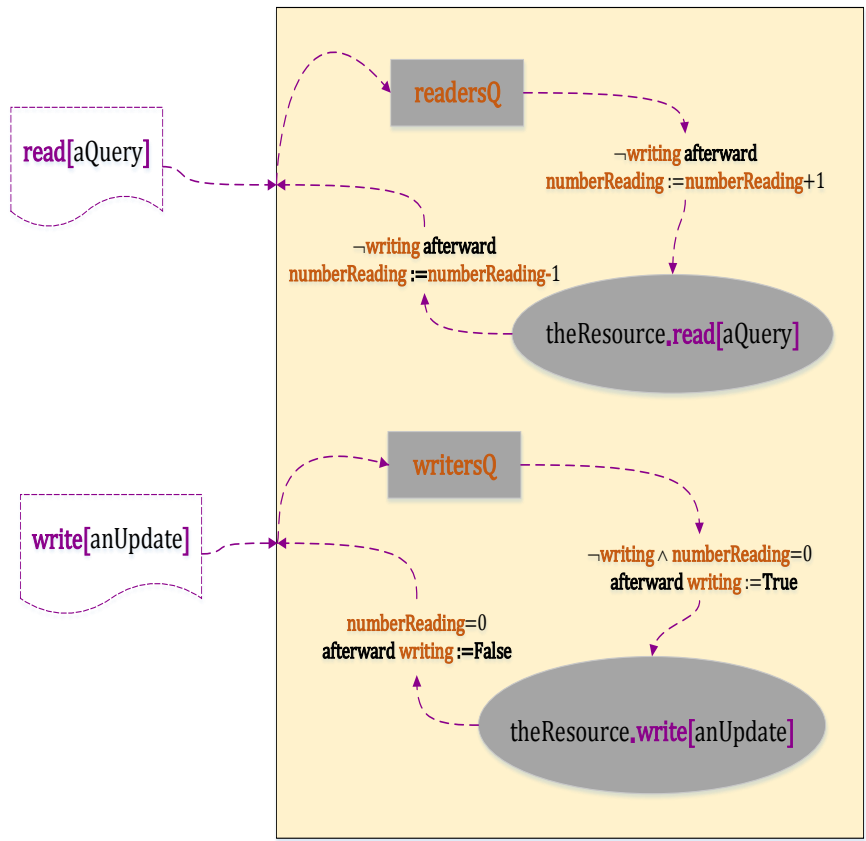

Note:
1. At most one activity is allowed to execute in the cheese.[ii]
2. The cheese has holes.[iii]
3. The value of a variable[iv] changes only when leaving the cheese or after an internal delegated operation.[v]

---

**Futures**

Futures [Baker and Hewitt 1977] are Actors that provide parallel execution.

Futures can be chained. For example,

> Size$_\bullet$[aFutureList:**Future**◁**List**◁**String**▷▷]:**Future**◁**Integer**▷ ≡
>   aFutureList ◈
>     **Future List**◁**String**▷[ ] ⸲
>       **Future** 0,
>     **Future List**◁**String**▷[aFirst:**String**,
>                       ∀aRest:**Future**◁**List**◁**String**▷▷] ⸲
>       **Future** aFirst$_\bullet$**length[ ]** + Size$_\bullet$[aRest] ⸮▮

The above procedure can compute the size of a list concurrently with creating the list.

**Future work**

As was the case with the lambda calculus and functional programming,[i] it has taken decades since they were invented [Hewitt, Bishop, and Steiger 1973] to understand the scientific and engineering of Actor Systems and it is still very much a work in progress.

Actors are becoming the default model of computation. C#, Java, JavaScript, Objective C, and SystemVerilog are all headed in the direction of the Actor Model and ActorScript is a natural extension of these languages. Since it is very close to practice, many programmers just naturally assume the Actor Model.

The following major developments in computer technology are pushing the Actor Model forward because Actor Systems are highly scalable:
- Many-core computer architectures
- Client-cloud computing

In fact, the Actor Model and ActorScript can be seen as codifying what are becoming some best programming practices for many-core and client-cloud computing.

**Conclusion**

The Actor Model is a mathematical theory that treats "*Actors*" as the universal primitives of concurrent digital computation. The model has been used both as a framework for a theoretical understanding of concurrency, and as the theoretical basis for several practical implementations of concurrent systems.

---

[i] For example, it took over four decades to develop the **eval** message-passing model of the lambda calculus [Hewitt, Bishop, and Steiger 1973, Hewitt 2011] building on the Lisp procedural model.

Unlike previous models of computation, the Actor Model was inspired by physical laws. It was also influenced by the programming languages Lisp, Simula 67 and Smalltalk-72, as well as ideas for Petri Nets, capability systems and packet switching. The advent of massive concurrency through client-cloud computing and many-core computer architectures has galvanized interest in the Actor Model.

When an Actor receives a message, it can concurrently:
- Send messages to (unforgeable) addresses of Actors that it has.
- Create new Actors.[i]
- Designate how to handle the next message it receives.

There is no assumed order to the above actions and they could be carried out concurrently. In addition two messages sent concurrently can be received in either order. Decoupling the sender from communication it sends was a fundamental advance of the Actor Model enabling asynchronous communication and control structures as patterns of passing messages.

Preferred methods for characterizing the Actor Model are as follows:
- *Axiomatically* stating laws that apply to all Actor *systems* [Baker and Hewitt 1977]
- *Denotationally* using the Computational Representation Theorem to characterize Actor computations [Clinger 1981; Hewitt 2006].
- O*perationally* using a suitable Actor programming language*, e.g.,* ActorScript [Hewitt 2012] that specifies how Actors can be implemented.

The Actor Model can be used as a framework for modeling, understanding, and reasoning about, a wide range of concurrent systems. For example:
- Electronic mail (e-mail) can be modeled as an Actor system. Accounts are modeled as Actors and email addresses as Actor addresses.
- Web Services can be modeled with endpoints modeled as Actor addresses.
- Objects with locks (e.g. as in Java and C#) can be modeled as Actors.
- The Actor Model can be a computational foundation for Inconsistency Robustness

The Actor Model supports Organizational Programming that is based on authority and accountability in iOrgs [Hewitt 2008a] with the goal of becoming an effective readily understood approach for addressing scalability issues in Software Engineering. The paradigm takes its inspiration from human organizations. iOrgs provide a framework for addressing issues of hierarchy, authority, accountability, scalability, and robustness using methods that are analogous to human organizations. Because humans are very familiar with the principles, methods, and practices of human organizations, they can transfer this knowledge and experience to iOrgs. iOrgs achieve scalability by mirroring human organizational structure. For example an iOrg can have sub-organizations specialized by areas such as sales, production, and so forth. Authority is delegated down the organizational structure and when necessary issues are escalated upward. Authority requires accountability for its use including record keeping and periodic reports. Management is in large part the art of reconciling authority and accountability.

---

[i] with new addresses

Actor technology will see significant application for integrating all kinds of digital information for individuals, groups, and organizations so their information usefully links together.

Information integration needs to make use of the following information system principles:

- **Persistence**. *Information is collected and indexed.*
- **Concurrency**: *Work proceeds interactively and concurrently, overlapping in time.*
- **Quasi-commutativity**: *Information can be used regardless of whether it initiates new work or become relevant to ongoing work.*
- **Sponsorship**: *Sponsors provide resources for computation, i.e., processing, storage, and communications.*
- **Pluralism**: *Information is heterogeneous, overlapping and often inconsistent.*
- **Provenance**: *The provenance of information is carefully tracked and recorded*

The Actor Model is intended to provide a foundation for inconsistency robust information integration.

## Acknowledgements

Important contributions to the semantics of Actors have been made by: Gul Agha, Beppe Attardi, Henry Baker, Will Clinger, Irene Greif, Carl Manning, Ian Mason, Ugo Montanari, Maria Simi, Scott Smith, Carolyn Talcott, Prasanna Thati, and Aki Yonezawa.

Important contributions to the implementation of Actors have been made by: Gul Agha, Bill Athas, Russ Atkinson, Beppe Attardi, Henry Baker, Gerry Barber, Peter Bishop, Nanette Boden, Jean-Pierre Briot, Bill Dally, Blaine Garst, Peter de Jong, Jessie Dedecker, Ken Kahn, Rajesh Karmani, Henry Lieberman, Carl Manning, Mark S. Miller, Tom Reinhardt, Chuck Seitz, Amin Shali, Richard Steiger, Dan Theriault, Mario Tokoro, Darrell Woelk, and Carlos Varela.

Research on the Actor Model has been carried out at Caltech Computer Science, Kyoto University Tokoro Laboratory, MCC, MIT Artificial

Intelligence Laboratory, SRI, Stanford University, University of Illinois at Urbana-Champaign Open Systems Laboratory, Pierre and Marie Curie University (University of Paris 6), University of Pisa, University of Tokyo Yonezawa Laboratory and elsewhere.

The Actor Model is intended to provide a foundation for scalable inconsistency-robust information integration in privacy-friendly client-cloud computing [Hewitt 2009b].

## Appendix 1. Historical background[16]

The Actor Model builds on previous models of nondeterministic computation. Several models of nondeterministic computation were developed including the following:

### Concurrency versus Turing's Model

Turing's model of computation was intensely psychological.[17] [Sieg 2008] formalized it as follows:

- *Boundedness:* A computer can immediately recognize only a bounded number of configurations.
- *Locality:* A computer can change only immediately recognizable configurations.

In the above, computation is conceived as being carried out in a single place by a device that proceeds from one well-defined state to the next.

Computations are represented differently in Turing Machines and Actors:

1. *Turing Machine*: a computation can be represented as a global state that determines all information about the computation.[18] It can be nondeterministic as to which will be the next global state.
2. *Actors*: a computation can be represented as a configuration. Information about a configuration can be indeterminate.[i]

### Lambda calculus

The Lambda calculus was originally developed as part of a system for the foundations of logic [Church 1932-33]. However, the system was soon shown to be inconsistent. Subsequently, Church removed logical propositions from the system leaving a purely procedural lambda calculus [Church 1941].[19]

However, the semantics of the lambda calculus were expressed using variable substitution in which the values of parameters were substituted into the body of an invoked lambda expression. The substitution model is unsuitable for concurrency because it does not allow the capability of sharing of changing resources.

That Actors which behave like mathematical functions exactly correspond with those definable in the lambda calculus provides an intuitive justification for the rules of the lambda calculus:

- *Lambda identifiers*: each identifier is bound to the address of an Actor. The rules for free and bound identifiers correspond to the Actor rules for addresses.

---

[i] For example, there can be messages in transit that will be delivered at some indefinite time.

- *Beta reduction*: each beta reduction corresponds to an Actor receiving a message. Instead of performing substitution, an Actor receives addresses of its arguments.

Inspired by the lambda calculus, the interpreter for the programming language Lisp [McCarthy *et. al.* 1962] made use of a data structure called an environment so that the values of parameters did not have to be substituted into the body of an invoked lambda expression.[20]

Note that in the definition in ActorScript [Hewitt 2011] of the lambda calculus below:
- All operations are local.
- The definition is modular in that each lambda calculus programming language construct is an Actor.
- The definition is easily extensible since it is easy to add additional programming language constructs.
- The definition is easily operationalized into efficient concurrent implementations.
- The definition easily fits into more general concurrent computational frameworks for many-core and distributed computation

The lambda calculus can be implemented in ActorScript as follows:

**Actor** thisIdentifier **Identifier**◁**aType**▷**[ ]**
                       // thisIdentifier is bound to this identifier
  **implements** **Expression**◁**aType**▷ **using**
    **eval**[anEnvironment]→ anEnvironment▪**lookup**[thisIdentifier]

**Actor** **ProcedureCall**◁**aType, AnotherType**▷
      [operator:(**[aType]**↦ **anotherType**), operand:**aType**]
  **implements** **Expression**◁**anotherType**▷ **using**
    **eval**[anEnvironment]→
     (operator▪**eval**[anEnvironment])▪**[**operand▪**eval**[environment]**]**

**Actor** **Lambda**◁**aType, AnotherType**▷
     [anIdentifier:**Identifier**◁**aType**▷, body:**anotherType**]
  **implements** **Expression**◁**[aType`]**↦ **anotherType**▷ **using**
    **eval**[anEnvironment]→
     [anArgument:**aType**]→
      body▪**eval**[**Environment**[anIdentifier,
                // create a new environment with **anIdentifier** bound to
                        anArgument,         // anArgument in
                        anEnvironment]**]**    // anEnvironment

**In many practical applications, simulating an Actor system using a lambda expression (*i.e.* using purely functional programming) is exponentially slower.**[21]

## Petri nets

Prior to the development of the Actor Model, Petri nets[22] were widely used to model nondeterministic computation. However, they were widely acknowledged to have an important limitation: they modeled control flow but not data flow. Consequently they were not readily composable thereby limiting their modularity.

Hewitt pointed out another difficulty with Petri nets:

> Simultaneous action, *i.e.,* the atomic step of computation in Petri nets is a transition in which tokens simultaneously disappear from the input places of a transition and appear in the output places. The physical basis of using a primitive with this kind of simultaneity seemed questionable to him.

Despite these apparent difficulties, Petri nets continue to be a popular approach to modeling nondeterminism, and are still the subject of active research.

## Capability Actor Systems

Capabilities were proposed in order to provide finer grained protection in operating systems [Dennis and van Horn 1966]. Unfortunately, capabilities have been awkward to use because their addresses were allocated in private memory of operating systems. The situation was considerably clarified by the development of the Actor Model in 1972. Unfortunately, the terms "capability" and "capability system" lacked axiomatizations and denotational semantics. Consequently, the terms were used in ambiguous and inconsistent ways. Capability systems can be considered to be approaches to security making use of specified principles that must include the laws of the Actor Model.

Capabilities were further developed in [Organick 1983, Levy 1984, Shapiro and Adams 2007, Woodruff, *et. al.* 2014]. Unfortunately, capabilities have been awkward to use because their addresses were allocated in private memory of operating systems. [Kwon, *et. al* 2014] is a tagged capability architecture that includes a special register to hold capabilities for addresses. The Object Capability Model [Miller 2006] has recommendations about best practices for implementing Actor systems.

Generally speaking, a capability is a token that contains an Actor address along with other information that can be used in sending messages to the Actor. The following are examples of capabilities:

- Waterken:[23] an Actor address of type **WebKey**
- Zebra Copy:[24] an Actor address together with additional information that includes a list of allowed message types

Capabilities were critiqued in [Rajunas 1989; Miller, Yee, and Shapiro 2003] concerning the following issues:

- *revocability*: Using proxies for Actors enables revocability  because messages are forwarded and so a proxy can revoke. Also revocation can be performed by communicating directly with an Actor.
- *multi-level security*:  Actors, *per se*, do *not* have levels of security although various security schemes can be implemented.[25]
- *delegation*:[26]  Actors[27] directly support delegation by passing addresses of Actors in messages.

[Miller 2006] followed up with the following analysis:

> *Just as we should not expect a base programming language to provide us all the data types we need for computation, we should not expect a base protection system to provide us all the elements we need to directly express access control policies. Both issues deserve the same kind of answer: We use the base to build abstractions, extending the vocabulary we use to express our solutions. In evaluating an access control model, one must examine how well it supports the extension of its own expressiveness by abstraction and composition.*

## Simula

Simula 1 [Nygaard 1962] pioneered nondeterministic discrete event simulation using a global clock:

> *In this early version of Simula a system was modeled by a (fixed) number of "stations", each with a queue of "customers". The stations were the active parts, and each was controlled by a program that could "input" a customer from the station's queue, update variables (global, local in station, and local in customer), and transfer the customer to the queue of another station. Stations could discard customers by not transferring them to another queue, and could generate new customers. They could also wait a given period (in simulated time) before starting the next action. Custom types were declared as data records, without any actions (or procedures) of their own.* [Krogdahl 2003]

Thus at each time step, the program of the next station to be simulated would update the variables.

Kristen Nygaard and Ole-Johan Dahl developed the idea (first described in an IFIP workshop in 1967) of organizing objects into "classes" with "subclasses" that could inherit methods for performing operations from their super classes. In this way, Simula 67 considerably improved the modularity of nondeterministic discrete event simulations.

According to [Krogdahl 2003]:

> *Objects could act as processes that can execute in "quasi-parallel" that is in fact a form of nondeterministic sequential execution in which a simulation is organized as "independent" processes. Classes in Simula 67 have their own procedures that start when an object is generated. However, unlike Algol procedures, objects may choose to temporarily stop their execution and transfer the control to another process. If the control is later given back to the object, it will resume execution where the control last left off. A process will always retain the execution control until it explicitly gives it away. When the execution of an object reaches the end of its statements, it will become "terminated", and can no longer be resumed (but local data and local procedures can still be accessed from outside the object).*

> *The quasi-parallel sequencing is essential for the simulation mechanism. Roughly speaking, it works as follows: When a process has finished the actions to be performed at a certain point in simulated time, it decides when (again in simulated time) it wants the control back, and stores this in a local "next-event-time" variable. It then gives the control to a central "time-manager", which finds the process that is to execute next (the one with the smallest next-event-time), updates the global time variable accordingly, and gives the control to that process.*

> *The idea of this mechanism was to invite the programmer of a simulation program to model the underlying system by a set of processes, each describing some natural sequence of events in that system (e.g. the sequence of events experienced by one car in a traffic simulation).*

> *Note that a process may transfer control to another process even if it is currently inside one or more procedure calls. Thus, each quasi-parallel process will have its own stack of procedure calls, and if it is not executing, its "reactivation point" will reside in the innermost of these calls. Quasi-parallel sequencing is analogous to the notion of co-routines [Conway 1963].*

Note that Simula operated on the global state of a simulation and not just on the local variables of simulated objects.[28] Also Simula-67 lacked formal interfaces and instead relied on inheritance in a hierarchy of objects thereby placing limitations to the ability to define and invoke behavior no directly inherited.

Types in Simula are the names of implementations called "classes" in contrast with ActorScript in which types are interfaces that do not name their implementation. Also, although Simula had nondeterminism, it did not have concurrency.[29]

## Planner

The two major paradigms for constructing semantic software systems were procedural and logical. The procedural paradigm was epitomized by using Lisp [McCarthy *et al.* 1962; Minsky, *et al.* 1968] recursive procedures operating on list structures. The logical paradigm was epitomized by uniform resolution theorem provers [Robinson 1965].

Planner [Hewitt 1969] was a kind of hybrid between the procedural and logical paradigms.[30] An implication of the form (P *implies* Q) was procedurally interpreted as follows:[31]

- **When asserted** P**, Assert** Q
- **When goal** Q**, SetGoal** P
- **When asserted** (*not* Q)**, Assert** (*not* P)
- **When goal** (*not* P)**, SetGoal** (*not* Q)

Planner was the first programming language based on the pattern-directed invocation of procedural plans from assertions and goals. ***It represented a rejection of the resolution uniform proof procedure paradigm.***

## Smalltalk-72

Planner, Simula 67, Smalltalk-72 [Kay 1975; Ingalls 1983] and packet-switched networks had previously used message passing. However, they were too complicated to use as the foundation for a mathematical theory of computation. Also they did not address fundamental issues of concurrency.

Alan Kay was influenced by message passing in the pattern-directed invocation of Planner in developing Smalltalk-71. Hewitt was intrigued by Smalltalk-71 but was put off by the complexity of communication that included invocations with many fields including global, sender, receiver, reply-style, status, reply, operator, *etc.*

In November 1972, Kay visited MIT and presented a lecture on some of his ideas for Smalltalk-72 building on the Logo work of Seymour Papert and the "little person" metaphor of computation used for teaching children to program. Smalltalk-72 made important advances in graphical user interfaces.

However, the message passing of Smalltalk-72 was quite complex [Kay 1975]. Code in the language was viewed by the interpreter as simply a stream of tokens. According to [Ingalls 1983]:[32]

> *The first (token) encountered (in a program) was looked up in the dynamic context, to determine the receiver of the subsequent message. The name lookup began with the class dictionary of the current activation. Failing there, it moved to the sender of that activation and so on up the sender chain. When a binding was finally found for the token, its value became the receiver of a new message, and the interpreter activated the code for that object's class.[33]*

Thus the message passing model in Smalltalk-72 was closely tied to a particular machine model and programming language syntax that did not lend itself to concurrency. Also, although the system was bootstrapped on itself, the language constructs were not formally defined as objects that respond to *eval* messages as in the definition of ActorScript [Hewitt 2010a].

## Actors

The invention of digital computers caused a decisive paradigm shift when the notion of an interrupt was invented so that input that is received asynchronously from outside could be incorporated in an ongoing computation. At first concurrency was conceived using low level machine implementation concepts like threads, locks, coherent memory, channels, cores, queues, *etc.*

The Actor Model [Hewitt, Bishop, and Steiger 1973; *etc.*] was based on message passing that was different from previous models of computation because the sender of a message is not intrinsic to the semantics of a communication.[34]

In contrast to previous global state model, computation in the Actor Model is conceived as distributed in space where computational devices called Actors communicate asynchronously using addresses of Actors and the entire computation is not in any well-defined state.[35] The local state of a serialized Actor is defined when it receives a message and at other times may be indeterminate.

Axioms of locality including *Organizational* and *Operational* hold as follows:

- *Organization:* The local storage of an Actor can include *addresses* only
  1. that were provided when it was created
  2. that have been received in messages
  3. that are for Actors created here
- *Operation:* In response to a message received, an Actor can
  1. create more Actors
  2. send messages[i] to *addresses* in the following:
     - the message it has just received
     - its local storage
  3. update its local storage for the next message

In concrete terms for Actor systems, typically we cannot observe the details by which the order in which an Actor processes messages has been determined. Attempting to do so affects the results. Instead of observing the internals of arbitration processes of Actor computations, we await outcomes.[36] Indeterminacy in arbiters produces indeterminacy in Actors.[ii]

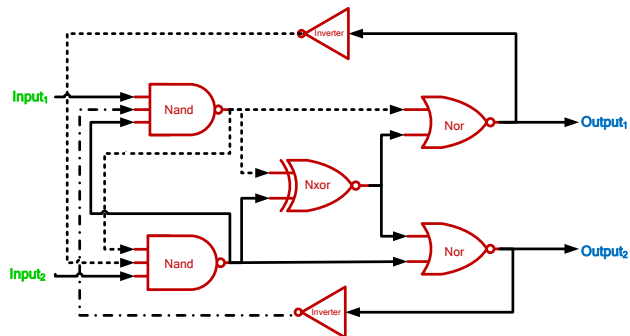

**Arbiter Concurrency Primitive**[37]

After the above circuit is started, it can remain in a meta-stable state for an unbounded period of time before it finally asserts either Output₁ or Output₂. So there is an inconsistency between the nondeterministic state model of computation and the circuit model of arbiters.[38]

The internal processes of arbiters are not public processes. Attempting to observe them affects their outcomes. Instead of observing the internals of arbitration processes, we necessarily await outcomes. Indeterminacy in arbiters

---

[i] Likewise the messages sent can contain addresses only
  1. that were provided when the Actor was created
  2. that have been received in messages
  3. that are for Actors created here

[ii] The dashed lines are used only to disambiguate crossing wires.

produces indeterminacy in Actors. The reason that we await outcomes is that we have no realistic alternative.

The Actor Model integrated the following:
- the lambda calculus
- interrupts
- blocking method invocation
- imperative programming using locks
- capability systems
- co-routines
- packet networks
- email systems
- Petri nets
- Smalltalk-72
- Simula-67
-  pattern-directed invocation (from Planner)

In 1975, Irene Greif published the first operational model of Actors in her dissertation. Two years after Greif published her operational model, Carl Hewitt and Henry Baker published the Laws for Actors [Baker and Hewitt 1977].

## Indeterminacy in Concurrent Computation

The first models of computation (*e.g.* Turing machines, Post productions, the lambda calculus, *etc.*) were based on mathematics and made use of a global state to represent a computational *step* (later generalized in [McCarthy and Hayes 1969] and [Dijkstra 1976]). Each computational step was from one global state of the computation to the next global state. The global state approach was continued in automata theory for finite state machines and push down stack machines, including their nondeterministic versions.[39] Such nondeterministic automata have the property of bounded nondeterminism; that is, if a machine always halts when started in its initial state, then there is a bound on the number of states in which it halts.[40]

Gordon Plotkin [1976] gave an informal proof as follows:

*Now the set of initial segments of execution sequences of a given nondeterministic program* P, *starting from a given state, will form a tree. The branching points will correspond to the choice points in the program. Since there are always only finitely many alternatives at each choice point, the branching factor of the tree is always finite.[41] That is, the tree is finitary. Now König's lemma says that if every branch of a finitary tree is finite, then so is the tree itself. In the present case this means that if every execution sequence of* P *terminates, then there are only finitely many execution sequences. So if an output set of* P *is infinite, it must contain a nonterminating computation.[42]*

The above proof is quite general and applies to the Abstract State Machine (ASM) model [Blass, Gurevich, Rosenzweig, and Rossman 2007a, 2007b; Glausch and Reisig 2006], which consequently are not really models of concurrency.

By contrast, the following Actor system can compute an integer of unbounded size using ActorScript™ [Hewitt 2010a]:[43]

```
Unbounded ≡
  start[ ]→                          // a start message is implemented by
    Let aCounter ← Counter[ ]        // let aCounter be a new Counter
      Do ⑪aCounter.go[ ] ⌐           // send aCounter a go message and concurrently
        ⑪aCounter.stop[ ]
                          // return the value of sending aCounter a stop message

Actor thisCounter Counter[ ]         // thisCounter is the name of this Actor
  count:= 0                          // the variable count is initially 0
  continue:= True
  stop[ ]→ count                              // return count
          afterward continue:=False
                          // continue is false for the next message received
  go[ ]→ continue ◆
        True ⸮         // if continue is True,
          Hole thisCounter.go[ ]      // send go[ ] to thisCounter after
            after count:=count+1      // incrementing count
        False ⸮ Void     // if continue is False, return Void
```

By the semantics of the Actor Model of computation [Clinger 1981] [Hewitt 2006], sending Unbounded a start message will result in return an integer of unbounded size.

***Theorem.*** There are nondeterministic computable functions on integers that cannot be implemented by a nondeterministic Turing machine.

   *Proof.* The above Actor system implements a nondeterministic function[i] that cannot be implemented by a nondeterministic Turing machine.

In many practical applications, simulating an Actor system using a Turing machine is exponentially slower.[44]

**Nondeterminism is a special case of Indeterminism.**

Consider the following Nondeterministic Turing Machine that starts at *Step 1*:
   *Step 1*:  Either print 1 on the next square of tape or execute *Step 3*.
   *Step 2*:  Execute *Step 1*.
   *Step 3*:  Halt.
According to the definition of Nondeterministic Turing Machines, the above machine might never halt.

---

[i] with graph {start[ ]⤳0, start[ ]⤳1, start[ ]⤳2, ...}

Note that the computations performed by the above machine are structurally different than the computations performed by the above Actor *counter* in the following way:

1. The decision making of the above Nondeterministic Turing Machine is internal (having an essentially psychological basis).
2. The decision making of the above Actor *Counter* exhibits physical indeterminacy.

Edsger Dijkstra further developed the nondeterministic global state approach, which gave rise to a controversy concerning *unbounded nondeterminism*[i]. Unbounded nondeterminism is a property of concurrency by which the amount of delay in servicing a request can become unbounded as a result of arbitration of contention for shared resources *while providing a guarantee that the request will be serviced*. The Actor Model provides the guarantee of service. In Dijkstra's model, although there could be an unbounded amount of time between the execution of sequential instructions on a computer, a (parallel) program that started out in a well-defined state could terminate in only a bounded number of states [Dijkstra 1976]. He believed that it was impossible to implement unbounded nondeterminism.

**Computation is not subsumed by logical deduction**

Kowalski claims that "*computation could be subsumed by deduction*"[45] The gauntlet was officially thrown in *The Challenge of Open Systems* [Hewitt 1985] to which [Kowalski 1988b] replied in *Logic-Based Open Systems*.[ii] This was followed up with [Hewitt and Agha 1988] in the context of the Japanese Fifth Generation Project.

According to Hewitt, *et. al.* and contrary to Kowalski computation in general cannot be subsumed by deduction and contrary to the quotation (above) attributed to Hayes computation in general is not subsumed by deduction. Hewitt and Agha [1991] and other published work argued that mathematical models of concurrency did not determine particular concurrent computations because they make use of arbitration for determining the order in which messages are processed. These orderings cannot be deduced from prior

---

[i] A system is defined to have *unbounded nondeterminism* exactly when both of the following hold:
1. When started, the system always halts.
2. For every integer n, the system can halt with an output that is greater than n.

[ii] [Kowalski 1979] forcefully stated:

*There is only one language suitable for representing information -- whether declarative or procedural -- and that is first-order predicate logic. There is only one intelligent way to process information -- and that is by applying deductive inference methods.*

information by mathematical logic alone. Therefore mathematical logic cannot implement concurrent computation in open systems.

A nondeterministic system is defined to have "*unbounded nondeterminism*"[i] exactly when both of the following hold:

1. When started, the system *always* halts.
2. For every integer n, it is possible for the system to halt with output that is greater than n.

This article has discussed the following points about unbounded nondeterminism controversy:

- A Nondeterministic Turing Machine cannot implement unbounded nondeterminism.
- A Logic Program[46] cannot implement unbounded nondeterminism.
- Semantics of unbounded nondeterminism are required to prove that a server provides service to every client.[47]
- An Actor system [Hewitt, *et. al*. 1973] can implement servers that provide service to every client and consequently unbounded nondeterminism.
- Dijkstra believed that unbounded nondeterminism cannot be implemented [Dijkstra 1967; Dijkstra and van Gasteren 1986].
- The semantics of CSP [Francez, Hoare, Lehmann, and de Roever 1979] specified bounded nondeterminism for reasons mentioned above in the article. Since Hoare *et. al.* wanted to be able to prove that a server provided service to clients, the semantics of a subsequent version of CSP were switched from bounded to unbounded nondeterminism.
- Unbounded nondeterminism was but a symptom of deeper underlying issues with sequential processes using nondeterministic global states as a foundation for computation.[ii]

The Computational Representation Theorem [Clinger 1981, Hewitt 2006] characterizes the semantics of Actor Systems without making use of sequential processes.

**Actor Model versus Classical Object Models**
The following are fundamental differences between Classical Object Models[Nygaard and Dahl 1967] and the Actor Model:

- Classical Object Models[48] are founded "a physical model, simulating the behavior of either a real or imaginary part of the world"[49], whereas the Actor Model is founded on the physics of computation.

---

[i] For example the following systems do *not* have unbounded nondeterminism:
- A nondeterministic system which sometimes halts and sometimes doesn't
- A nondeterministic system that always halts with an output less than 100,000.
- An operating system that never halts.

[ii] See [Knabe 1992].

- Every Classical Object is an instance of a Class in a hierarchy[50], whereas an Actor can implement multiple interfaces.[51]
- Virtual Procedures can be used to operate on Objects, whereas messages[i] can be sent to Actors.[52]

**Hairy Control Structure**

Peter Landin introduced a powerful co-routine control structure using his **J** (for Jump) operator that could perform a nonlocal goto into the middle of a procedure invocation [Landin 1965]. In fact the **J** operator enabled a program to jump back into the middle of a procedure invocation even after it had already returned!

[Reynolds 1972] introduced control structure continuations using a primitive called **escape** that is a more structured versions of Landin's **J** operator. Sussman and Steele called their variant of **escape** by the name "*call with current continuation*." General use of **escape** is not compatible with usual stack disciple introducing considerable operational inefficiency. Also, using *escape* can leave customers stranded. Consequently, use of **escape** is generally avoided these days and exceptions[53] are used instead so that clean up can be performed.

In the 1960's at the MIT AI Lab a remarkable culture grew up around "*hacking*" that concentrated on remarkable feats of programming.[54] Growing out of this tradition, Gerry Sussman and Guy Steele decided to try to understand Actors by reducing them to machine code that they could understand and so developed a "*Lisp-like language, Scheme, based on the lambda calculus, but extended for side effects, multiprocessing, and process synchronization*." [Sussman and Steele 1975].[55]

Their reductionist approach included primitives like the following:
- START!PROCESS
- STOP!PROCESS
- EVALUATE!UNINTERRUPTIBLEY[ii]

that had the following explanation:[56]

Of course, the above reductionist approach is very unsatisfactory because it missed a crucial aspect of the Actor Model: *the reception ordering of message*s.

---

[i] A message can be one-way and each must be of type **Message**.

[ii] "*This is the synchronization primitive. It evaluates an expression uninterruptedly; i.e. no other process may run until the expression has returned a value.*"

McDermott, and Sussman [1972] developed the Lisp-based language Conniver based on "hairy control structure" that could implement non-chronological backtracking that was more general than the chronological backtracking in Planner. However, Hewitt and others were skeptical of hairy control structure.

Pat Hayes remarked:

> *Their* [Sussman and McDermott] *solution, to give the user access to the implementation primitives of Planner, is however, something of a retrograde step (what are Conniver's semantics?).* [Hayes 1974]

Hewitt had concluded:

> *One of the most important results that has emerged from the development of Actor semantics has been the further development of techniques to semantically analyze or synthesize control structures as patterns of passing messages. **As a result of this work, we have found that we can do without the paraphernalia of "hairy control structure."*** [57] (emphasis in original)

Sussman and Steele [1975] noticed some similarities between Actor programs and the lambda calculus. They mistakenly concluded that they had reduced Actor programs to a "continuation-passing programming style":

> *It is always possible, if we are willing to specify explicitly what to do with the answer, to perform any calculation in this way: rather than reducing to its value, it reduces to an application of a continuation to its value. That is, in this continuation-passing programming style, **a function always "returns" its result by "sending" it to another function**.* (emphasis in original)

However, some Actor programming language constructs are not reducible to a continuation-passing style. For example, futures are not reducible to continuation-passing style.

On the basis of their experience, Sussman and Steele developed the general thesis that Actors were merely the lambda calculus in disguise. Steele [1976] in the section "Actors ≡ Closures (mod Syntax)" disagreed with Hewitt who had "*expressed doubt as to whether these underlying continuations can themselves be expressed as lambda expressions.*" However, customers cannot be expressed as lambda expressions because doing so would preclude being able to enforce the requirement that a customer will process at most one response (*i.e.* exception or value return). Also implementing customers as lambda expressions can leave customers stranded.

In summary, Sussman and Steele [1975] mistakenly concluded "*we discovered that the 'Actors' and the lambda expressions were identical in implementation.*"[58] The actual situation is that the lambda calculus is capable of expressing some kinds of sequential and parallel control structures but, in general, *not* the concurrency expressed in the Actor Model.[59] On the other hand, the Actor Model is capable of expressing everything in the lambda calculus [Hewitt 2008f] and is exponentially faster for important applications like information integration [Hewitt 2012].

For example, futures can be adaptively created to do the kind of computation performed by hairy structure. [Hewitt 1974] invented the same-fringe problem as an illustration where the "fringe" of a tree is a list of all the leaf nodes of the tree.

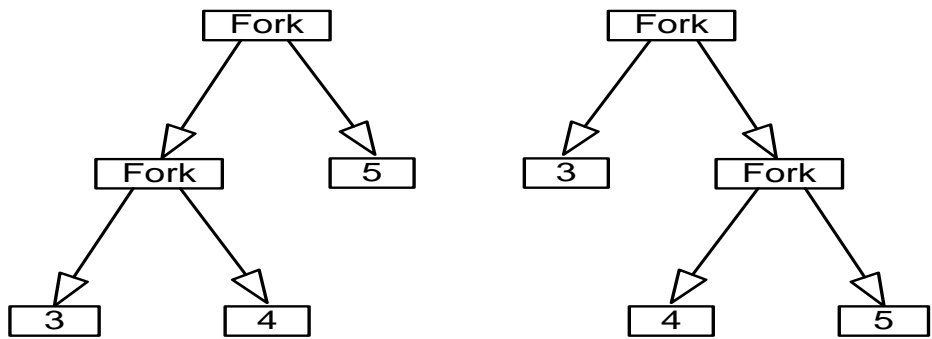

**Two trees with the same fringe [3 4 5]**

SameFringe▪[first:**Tree**, second:**Tree**] ≡
    Fringe▪[first] = Fringe▪[second]▮

Fringe▪[aTree:**Tree**] ≡    // the fringe of a**Tree** of type **Tree** is defined to be
    aTree ◆Leaf[x] ⦂ [x]    // the fringe of a leaf is a list with just its contents
        Fork[left, right] ⦂        // the fringe of a fork is
            [⩒Fringe▪[left],        // the fringe of its left branch
            ⩒**Postpone** Fringe▪[right]]▮
                  // appended to the fringe of its right branch

Using Actors in this way obviates the need for explicit co-routine primitives, *e.g.*, *yield* in C# [ECMA 2006], JavaScript [ECMA 2014], *etc*.

ActorScript makes use of a variant of "continuation passing style" called "string bean style" [Hewitt 2011] in which continuations are not made explicit while programs are required to be linear between holes in the cheese.

### Early Actor Programming languages

Henry Lieberman, Dan Theriault, *et al.* developed Act1, an Actor programming language. Subsequently for his master's thesis, Dan Theriault developed Act2. These early proof of concept languages were rather inefficient and not suitable for applications. In his doctoral dissertation, Ken Kahn developed Ani, which he used to develop several animations. Bill Kornfeld developed the Ether programming language for the Scientific Community Metaphor in his doctoral dissertation. William Athas and Nanette Boden [1988] developed Cantor which is an Actor programming language for scientific computing. Jean-Pierre Briot [1988, 1999] developed means to extend Smalltalk 80 for Actor computations. Darrell Woelk [1995] at MCC developed an Actor programming language for InfoSleuth agents in Rosette.

Hewitt, Attardi, and Lieberman [1979] developed proposals for delegation in message passing. This gave rise to the so-called inheritance anomaly controversy in concurrent programming languages [Satoshi Matsuoka and Aki Yonezawa 1993, Giuseppe Milicia and Vladimiro Sassone 2004]. ActorScript [Hewitt 2010] has proposal for addressing delegation issues.

## Garbage Collection

Garbage collection (the automated reclamation of unused storage) was an important theme in the development of the Actor Model.

In his doctoral dissertation, Peter Bishop developed an algorithm for garbage collection in distributed systems. Each system kept lists of links of pointers to and from other systems. Cyclic structures were collected by incrementally migrating Actors (objects) onto other systems which had their addresses until a cyclic structure was entirely contained in a single system where the garbage collector could recover the storage.

Henry Baker developed an algorithm for real-time garbage collection is his doctoral dissertation. The fundamental idea was to interleave collection activity with construction activity so that there would not have to be long pauses while collection takes place.

Lieberman and Hewitt [1983] developed a real time garbage collection based on the lifetimes of Actors (Objects). The fundamental idea was to allocate Actors (objects) in generations so that only the latest generations would have to be examined during a garbage collection.

## Cosmic Cube

The Cosmic Cube was developed by Chuck Seitz *et al.* at Caltech providing architectural support for Actor systems. A significant difference between the Cosmic Cube and most other parallel processors is that this multiple instruction multiple-data machine used message passing instead of shared variables for communication between concurrent processes. This computational model was reflected in the hardware structure and operating system, and also the explicit message passing communication seen by the programmer.

## Communicating Sequential Processes

Arguably, the first concurrent programs were interrupt handlers. During the course of its normal operation, a computer needed to be able to receive information from outside (characters from a keyboard, packets from a network, *etc.*). So when the information was received, execution of the computer was "interrupted" and special code called an interrupt handler was called to *put* the information in a buffer where it could be subsequently retrieved.

In the early 1960s, interrupts began to be used to simulate the concurrent execution of several programs on a single processor. Having concurrency with shared memory gave rise to the problem of concurrency control. Originally, this problem was conceived as being one of mutual exclusion on a single computer. Edsger Dijkstra developed semaphores and later, [Hoare 1974, Brinch Hansen 1996] developed monitors to solve the mutual exclusion problem. However, neither of these solutions provided a programming language construct that encapsulated access to shared resources. This problem was remedied by the introduction of serializers [Hewitt and Atkinson 1977, 1979; Atkinson 1980].

His belief was manifested in his theory of computation based on "weakest preconditions" for global states of computation [Dijkstra 1976]. He argued that unbounded nondeterminism results in non-continuity of his weakest precondition semantics. In sum, Dijkstra was certain that unbounded nondeterminism is impossible to implement.

Hoare was convinced by Dikstra's argument. Consequently, the semantics of CSP specified bounded nondeterminism.

Consider the following program written in CSP [Hoare 1978]:

```
[X :: Z!stop( )                        ① In process X, send Z a stop message
 ||                                    ① process X operates in parallel with process Y
Y :: guard: boolean; guard := true;
                                       ① In process Y, initialize boolean variable guard to true and then
     *[guard→ Z!go( ); Z?guard]
                                       ① while guard is true, send Z a go message and then input guard from Z
 ||                                    ① process Y operates in parallel with process Z
Z :: n: integer; n:= 0;  ① In process Z, initialize integer variable n to 0 and then
       continue: boolean; continue := true;
                                       ① initialize boolean variable continue to true and then
     *[                                ① repeatedly either
        X?stop( ) → continue := false;
                                       ① input a stop message from X, set continue to false and then
               Y!continue;             ① send Y the value of continue
        []                             ① or
        Y?go( )→ n := n+1;
                                       ① input a go message from Y, increment n, and then
               Y!continue]]            ① send Y the value of continue
```

According to Clinger [1981]:

    this program illustrates global nondeterminism, since the nondeterminism arises from incomplete specification of the timing of signals between the three processes X, Y, and Z. The repetitive guarded command in the definition of Z has two alternatives: either the stop message is accepted from X, in which case continue is set to false, or a go message is accepted from Y, in which case n is incremented and Y is sent the value of continue. If Z ever accepts the stop message from X, then X terminates. Accepting the stop causes continue to be set to false, so after Y sends its next go message, Y will receive false as the value of its guard and will terminate. When both X and Y have terminated, Z terminates because it no longer has live processes providing input.

    As the author of CSP points out, therefore, if the repetitive guarded command in the definition of Z were required to be fair, this program would have unbounded nondeterminism: it would be guaranteed to halt but there would be no bound on the final value of n. In actual fact, the repetitive guarded commands of CSP are not required to be fair, and so the program may not halt [Hoare 1978]. This fact may be confirmed by a tedious calculation using the semantics of CSP [Francez, Hoare, Lehmann, and de Roever 1979] or simply by noting that the semantics of CSP is based upon a conventional power domain and thus does not give rise to unbounded nondeterminism.

But Hoare knew that trouble was brewing in part because for several years, proponents of the Actor Model had been beating the drum for unbounded nondeterminism. To address this problem, he suggested that implementations of CSP should be as close as possible to unbounded nondeterminism! But his suggestion was difficult to achieve because of the nature of communication in CSP using nondeterministic select statements (from nondeterministic state machines, *e.g.*, [Dijkstra 1976]), which in the above program which takes the form

> [X?stop( ) → ...
>   []
> Y?go( ) → ...]

The structure of CSP is fundamentally at odds with guarantee of service.

Using the above semantics for CSP, it was impossible to formally prove that a server actually provides service to multiple clients (as had been done previously in the Actor Model). That's why the semantics of CSP were reversed from bounded non-determinism [Hoare CSP 1978] to unbounded non-determinism [Hoare CSP 1985]. However, bounded non-determinism was but a symptom of deeper underlying issues with nondeterministic transitions in communicating sequential processes (see [Knabe 1992]).

## Smalltalk-80

Smalltalk-72 progressed to Smalltalk-80[Alan Kay, Dan Ingalls, Adele Goldberg, Ted Kaehler, Diana Merry, Scott Wallace, Peter Deutsch], which introduced the code browser as an important innovation. For example, the following diagram depicts a code-browser window:

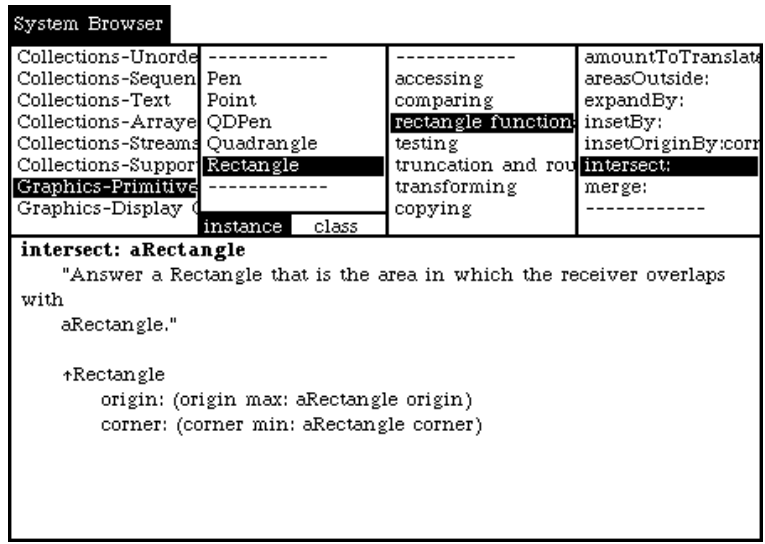

## π-Calculus Actors

Robin Milner's initial published work on concurrency [Milner 1973] was notable in that it was not overtly based on sequential processes, although computation still required sequential execution (see below).

His work differed from the previously developed Actor Model in the following ways:

- There are a fixed number of processes as opposed to the Actor Model which allows the number of Actors to vary dynamically
- The only quantities that can be passed in messages are integers and strings as opposed to the Actor Model which allows the addresses of Actors to be passed in messages
- The processes have a fixed topology as opposed to the Actor Model which allows varying topology
- Communication is synchronous as opposed to the Actor Model in which an unbounded time can elapse between sending and receiving a message.
- Unlike the Actor Model, there is no reception ordering and consequently there is only bounded nondeterminism. However, with bounded nondeterminism it is impossible to prove that a server guarantees service to its clients, *i.e.*, a client might starve.

Building on the Actor Model, Milner [1993] removed some of these restrictions in his work on the π-calculus:

*Now, the pure lambda-calculus is built with just two kinds of thing: terms and variables. Can we achieve the same economy for a process calculus? Carl Hewitt, with his Actors model, responded to this challenge long ago; he declared that a value, an operator on values, and a process should all be the same kind of thing: an Actor.*

*This goal impressed me, because it implies the homogeneity and completeness of expression ...*

*So, in the spirit of Hewitt, our first step is to demand that all things denoted by terms or accessed by names--values, registers, operators, processes, objects--are all of the same kind of thing....*

**However, some fundamental differences remain between the Actor Model and the π–calculus**

:

- The Actor Model is founded on physics whereas the π–calculus is founded on algebra.
- Semantics of the Actor Model is based on message orderings in the Computational Representation Theorem. Semantics of the π–calculus is based on structural congruence in various kinds of bi-simulations and equivalences.[60]

Communication in the π -calculus takes the following form:

- *input:* u[x].P is a process that gets a message from a communication channel u before proceeding as P, binding the message received to the identifier x. In ActorScript [Hewitt 2010a], this can be modeled as follows: **Let** x←u•**get[ ]**; P[61]

- *output:* ū[m].P is a process that puts a message m on communication channel u before proceeding as P. In ActorScript, this can be modeled as follows: **Do** u**.put[x]**; P[62]

The above operations of the π-calculus can be implemented in Actor systems using a two-phase commit protocol [Knabe 1992; Reppy, Russo, and Xiao 2009]. The overhead of communication in the π–calculus presents difficulties to its use in practical applications.

Process calculi (*e.g.* [Milner 1993; Cardelli and Gordon 1998]) are closely related to the Actor Model. There are similarities between the two approaches, but also many important differences (philosophical, mathematical and engineering):

- There is only one Actor Model (although it has numerous formal systems for design, analysis, verification, modeling, etc.) in contrast with a variety of species of process calculi.
- The Actor Model was inspired by the laws of physics and depends on them for its fundamental axioms in contrast with the process calculi being inspired by algebra [Milner 1993].
- Unlike the Actor Model, the sender is an intrinsic component of process calculi because they are defined in terms of reductions (as in the lambda calculus).
- Processes in the process calculi communicate by sending messages either through channels (synchronous or asynchronous), or via ambients (which can also be used to model channel-like communications [Cardelli and Gordon 1998]). In contrast, Actors communicate by sending messages to the addresses of other Actors (this style of communication can also be used to model channel-like communications using a two-phase commit protocol [Knabe 1992]).

There remains a Great Divide between process calculi and the Actor Model:

- *Process calculi:* algebraic equivalence, bi-simulation [Park 1980], *etc.*
- *Actor Model:* futures [Baker and Hewitt 1977], Swiss cheese, garbage collection, *etc.*

## J–Machine

The J–Machine was developed by Bill Dally *et al.* at MIT providing architectural support suitable for Actors. This included the following:

- Asynchronous messaging
- A uniform space of Actor addresses to which messages could be sent concurrently regardless of whether the recipient Actor was local or nonlocal
- A form of Actor pipelining

Concurrent Smalltalk (which can be modeled using Actors) was developed to program the J Machine.

**"Fog Cutter" Actors**

[Karmani and Agha 2011] promoted "*Fog Cutter*"[i] Actors each of which is required to have a mailbox, thread, state, and program diagrammed as follows:[63]

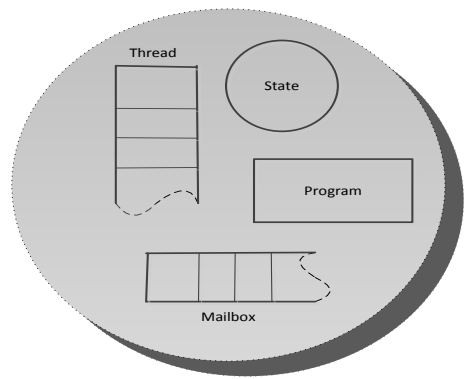

**Event loop: Process a message from the Mailbox using the Thread, then reset the Thread stack thereby completing the message-passing turn**

However, Fog Cutter Actors are special cases because:[ii]

- *Each Fog Cutter Actor has a 'mailbox'.* But if everything that interacts is an Actor, then a mailbox must be an Actor and so in turn needs a mailbox which in turn…[Hewitt, Bishop, and Steiger 1973] Of course, mailboxes having mailboxes is an infinite regress that has been humorously characterized by Erik Meijer as "down the rabbit hole." [Hewitt, Meijer, and Szyperski 2012]
- *A Fog Cutter Actor 'terminates' when every Actor that it has created is 'idle' and there is no way to send it a message.* In practice, it is preferable to use garbage collection for Actors that are inaccessible. [Baker and Hewitt 1977]
- *Each Fog Cutter Actor executes a 'loop' using its own sequential 'thread' that begins with receiving a message followed by possibly creating more Actors, sending messages, updating its local state, and then looping back for the next message to complete a 'turn'.* In practice, it is preferable to provide "Swiss cheese" by which an Actor can concurrently process

---

[i] so dubbed by Kristen Nygaard (private communication).
[ii] "Fog Cutter" is in *italics*.

multiple messages without the limitation of a sequential thread loop. [Hewitt and Atkinson 1977, 1979; Atkinson 1980; Hewitt 2011]

- *A Fog Cutter Actor has a well-defined local 'autonomous' 'state' that can be updated* [64] *while processing a message.* However, because of indeterminacy an Actor may not be in a well-defined local independent state. For example, Actors might be entangled[65] with each other so that their actions are correlated. Also, large distributed Actors (*e.g.* [www.dod.gov](www.dod.gov)) do not have a well-defined state. In practice, it is preferable for an Actor not to change its local information while it is processing a message and instead specify to the information to be used in how it will process the next message received (as in ActorScript [Hewitt 2011]).

Fog Cutter Actors have been extremely useful for exploring issues about Actors including the following alternatives:

- **Reception order of messaging** instead of *Mailbox*
- **Activation order of messaging** instead of *Thread*
- **Behavior** instead of *State+Program*

In practice, the most common and effective way to explain Actors has been *operationally* using a suitable Actor programming language (*e.g.,* ActorScript [Hewitt 2012]) that specifies how Actors can be implemented along with an English explanation of the axioms for Actors (*e.g.*, as presented in this paper).

Concurrency control for readers and writers in a shared resource is a classic problem. The fundamental constraint is that multiple writers are not allowed to operate concurrently and a writer is not allowed operate concurrently with a reader.

The interface for the readers/writer guardian is the same as the interface for the shared resource:

   **Interface ReadersWriter having read[Query]↦ QueryResult,**
   **write[Update]↦ Void**

State diagram of **ReadersWriter** implementations:

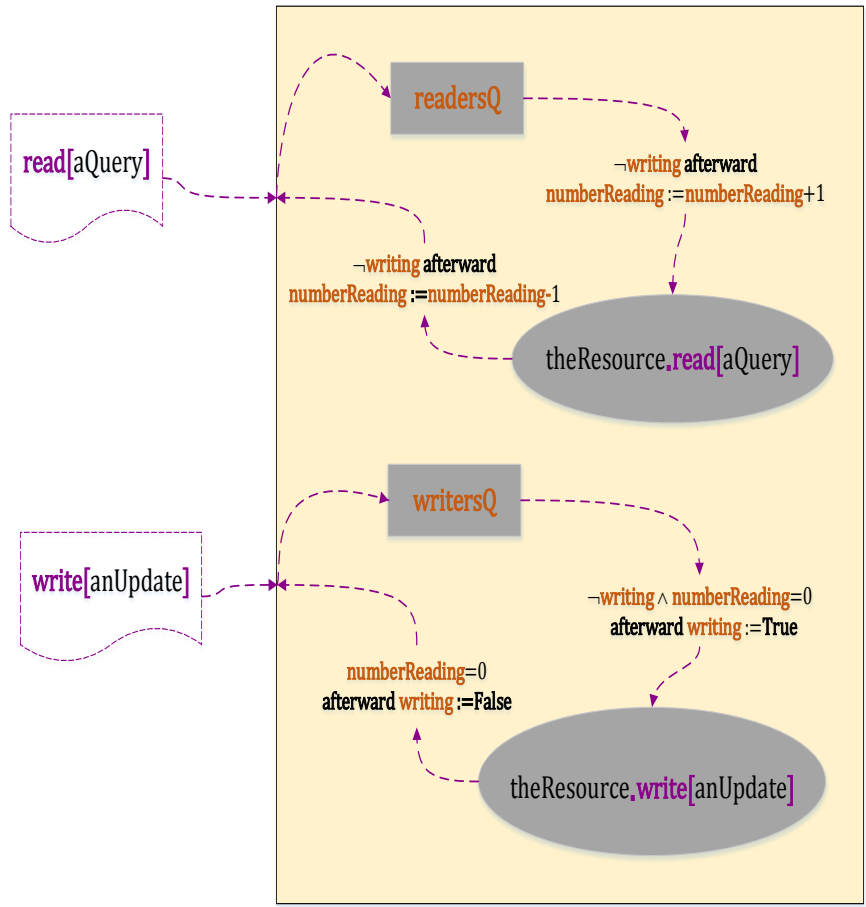

Note:
4. At most one activity is allowed to execute in the cheese.[i]
5. The cheese has holes.[ii]
6. The value of a variable[iii] can change only when leaving the cheese or after an internal delegated operation.[iv]

---

[i] Cheese is yellow in the diagram
[ii] A hole is grey in the diagram
[iii] A variable is orange in the diagram
[iv] Of course, other external Actors can change.

**Erlang Actors**

Erlang Actors [Armstrong 2010] are broadly similar to Fog Cutter Actors:
1. Each Erlang Actor is a process that does not share memory with other processes.
2. An Erlang Actor can retrieve a message from its mailbox by selectively removing a message matching a particular pattern.

However, Erlang Actors have the following issues:
- Erlang imposes the overhead that messages sent between two Erlang Actors are delivered in the order they are sent.
- Instead of using exception handling, Erlang Actors rely on process failure[i] propagating between processes and their spawned processes.
- Instead of using garbage collection to recover storage and processing of unreachable Actors, each Erlang Actor must perform an internal termination or be killed. However, data structures *within* a process are garbage collected.

Erlang Actors have been used in high-performance applications. For example, Ericsson uses Erlang in 3G mobile networks worldwide [Ekeroth and Hedström 2000].

**Sqeak**

Squeak [Ingalls, Kaehler, Maloney, Wallace, and Kay 1997] is a dialect of Smalltalk-80 with added mechanisms of islands, asynchronous messaging, players and costumes, language extensions, projects, and tile scripting. Its underlying object system is class-based, but the user interface is programmed as though it is prototype-based.

**Orleans Actors**

Orleans [Bykov, Geller, Kliot, Larus, Pandya, and Thelin 2010; Bernstein, Bykov, Geller, Kliot, and Thelin 2014] is a *distributed* implementation of Actors that transparently sends messages between Actors on different computers enabling greater scalability and reliability of practical applications.

Orleans is based on single-threaded Actor message invocations. An Actor processes a message using a thread from a thread pool. When the message has been processed, the thread can be returned to the thread pool.[66]

That an Orleans Actor does not share memory with other Actors is enforced by doing a deep copy of messages if required.

---

[i] based on an arbitrary time-out

A globally unique identifier[67] is created for each Orleans Actor with a consequence that there is extra storage overhead that can be significant for a very small Orleans Actor.[68] A globally unique identifier can be used to send a message, which will, if necessary, create an activation[69] of an Orleans Actor in the memory of a process.[70]

- Orleans allows the use of strings and long integers as globally unique identifiers in order to provide for perpetual Actors whose storage can only be collected using potentially unsafe means, which can result in a dangling globally unique identifier.
- A system design choice was made in Orleans not to use automated storage reclamation technology (garbage collection) to keep track of whether an Orleans Actor could have been forgotten by all applications and thus become inaccessible. Consequently, Orleans can have the following inefficiencies:
  - o A short-lived Orleans Actor that has become inaccessible *does not have its storage in the process quickly recycled* resulting in a larger working set and decreased locality of reference.[71]
  - o A long-lived Orleans Actor that has become inaccessible *does not ever have its storage recycled*[72] resulting in larger memory requirements.[73] However, collection of the storage of long-lived Actors is not so important in some applications because long-term memory has become relatively inexpensive.

An Orleans Actor ties up a thread while it is taking a turn to process a message regardless of the amount of time required, *e.g.*, time to make a system call. In this way, Orleans avoids timing races in the value of a variable of an Actor.[i] A consequence of being single-threaded can be reduced performance of Orleans Actors as follows:

- lack of parallelism in processing a message
- lack of concurrency between processing a message and executing waiting method calls invoked by processing the message.[74]
- thread-switching overhead between sending and receiving a message to an Orleans Actor in the same process[75]

---

[i] ActorScript goes even further in this direction by enforcing that an Actor can change the value of a variable only when it is leaving the cheese or after an internal delegated operation.

A waiting method call can be resolved using the **await**[76] primitive as follows:

    **await** *anActor.aMethodName*(…)[i]

For example:

    **var** *anActor = aFactory*.GetActor(*aGloballyUniqueIdentifier*);
    **try** {…*aUse*(**await** *anActor.aMethodName*(…))…
           *anotherUse*(**await** *anActor.anotherMethodName*(…))…}
     **catch** …;[ii]

When reentrancy[77] is enabled, the method calls for *aMethodName* and *anotherMethodName* above are executed *after* the current message-processing turn:

- If completed successfully, the value of a waiting method call is supplied in a *new* turn at the point of method invocation, *e.g.*, the value of the method call for *aMethodName* of is supplied to *aUse*.
- If a waiting method call throws an exception, it is given to the exception handler in a new turn.

Orleans uses C# compiler "stack ripping" to use behind-the-scenes sequential turns to execute waiting method calls.

A message sent to an Orleans Actor must return a promise[78] Actor[79], which is a version of a future Actor. A promise Actor for a method call *anActor.aMethodName*(…) can be created using the following code:[iii]

    **try** {**return** Task.FromResult(**await** *anActor.aMethodName*(…));}
     **catch** (Exception anException)
      {**return** Task.FromException(anException);}[iv]

Note that a promise is *not* an Orleans Actor because it does not have a globally unique identifier.[v]

One of the motivations for the requirement that Orleans Actors must return promises when sent messages is to enable the **await** primitive to *hide* promises so that clients of Orleans Actors do not have to deal with the

---

[i] ActorScript uses ↓*aFuture* to resolve *aFuture*

[ii] In ActorScript the program is:

    **Try** …*aUse*(ⓐ*anActor.aMethodName*(...))…
         *anotherUse*(ⓐ*anActor.anotherMethodName*(...))…
     **catch** …

[iii] ActorScript uses **Future** *anExpression* to create a future for *anExpression*

[iv] There is an inefficiency in the above code in that the method call returns a promise that is taken apart and then an equivalent promise is created to be returned.

[v] It would be impractical for promises to be Orleans Actors because

- they are created as the return value of *every* Orleans Actor method call
- the storage of Orleans Actors is not recovered

return type `Task<T>` of each Orleans Actor method call for some application type `T`.

Orleans is an important step in furthering a goal of the Actor Model that application programmers need not be so concerned with low-level system details.[i] For example, in moving to the current version, Orleans reinforces the current trend of not exposing customer Actors[80] to application programmers.[81]

As a research project, Orleans had to make some complicated tradeoffs to implement more reliable distributed Actors. Implementing Actor systems that are both *robust* and *performant* is an extremely challenging research project that has taken place over many decades. More research remains to be done. However, Orleans has already been used in some high-performance applications including multi-player computer games, *e.g.*, Halo[Bykov 2013].

### JavaScript Actors

JavaScript Actors are broadly similar to Fog Cutter Actors.[82]

A future version of JavaScript[83] will include an `await`[84] primitive that can be used to resolve promise[ii] Actors, which enables application programmers not to have to write so much "string-bean" continuation-passing code.[85]

For example, the following expression
```
   (↓Future aSlowActor.do[10, 20]) + ↓Future aSlowActor.do[30,40]
```
can be accomplished as follows:[iii]
```
   (await future(() => aSlowActor.do(10, 20)))
    + await future(() => aSlowActor.do(30, 40))
```
where
```
   function future(thunkForExpression)
         // a thunk is an intermediary procedure for assistance in carrying out a task
      {return Promise.resolve(true)
                     .then((aValueToDiscard) =>
                              thunkForExpression())};
```

---

[i] *e.g.* threads, throttling, load distribution, cores, persistence, automated storage reclamation, locks, location transparency, channels, ports, *etc.*

[ii] promise Actors were originally called "futures" in JavaScript

[iii] this expression must be *directly* inside an `async function.`

There is a potential pitfall in the use of JavaScript promises in that the following substitute code for the above does *not* work to *concurrently* execute the two calls to `aSlowActor`:[86]

```
(await
   new Promise((aPromiseValueSetter) =>
      // a promise-value setter[87] is a procedure that sets the value of a promise
                aPromiseValueSetter(aSlowActor.do(10, 20))))
 + await
     new Promise((aPromiseValueSetter) =>
                aPromiseValueSetter(aSlowActor.do(30, 40)))
```

Note that neither of the two promise-value setters in the above code is called more than once to set the value of a promise. However, the future version of JavaScript will make use[88] of the ability to call a promise-value setter multiple times. If a promise-value setter is called twice to set the value of a promise, an exception is *not* thrown. Instead, the second call fails *silently*. The future version of JavaScript will make use of asynchronous *races* in calling a promise-value setter. The *first* call to the promise-value setter *wins* and subsequent calls fail *silently*.

To implement parallelism, JavaScript has workers.[89] Although multiple workers can reside in a process, they do not share memory addresses and consequently cannot efficiently communicate using many-core coherency.[90] A worker communicates with other workers using blobs[i] in order to guarantee memory separation. Each worker acts as a *single-threaded*, *non-preemptive* time-sharing system for processing messages for Actors that reside in its memory. However, JavaScript workers have the following efficiency issues:[91]

1. There is no parallelism in processing messages for different Actors on a worker and the processing of a message by a slowly executing Actor *cannot* be preempted thereby bringing *all*[ii] other work on the worker to a *standstill*.[iii]

2. An Actor on a worker can communicate with Actors on other workers using message ports *only* by sending messages that are *blobs*.

3. Addresses of Actors on other workers must be *blobbed*.[92]

---

[i] a blob is a data structure that cannot contain pointers. In the past, a more limited meaning called <u>BLOB</u> has been used as an acronym for <u>B</u>inary <u>L</u>arge <u>OB</u>ject.

[ii] including any queued promises

[iii] Issues of non-preemption motivated the invention of time-slicing [Bemer 1957] by which tasks are switched at the expiration of a timer.

## Was the Actor Model premature?

The history of the Actor Model raises the question of whether it was premature.

### Original definition of prematurity

As originally defined by [Stent 1972], "A discovery is premature if its implications cannot be connected by a series of simple logical steps to contemporary canonical or generally accepted knowledge." [Lövy 2002] glossed the phrase "series of simple logical steps" in Stent's definition as referring to the "*target community's ways of asking relevant questions, of producing experimental results, and of examining new evidence.*" [Ghiselin 2002] argued that if a "*minority of scientists accept a discovery, or even pay serious attention to it, then the discovery is not altogether premature in the Stentian sense.*" In accord with Ghiselin's argument, the Actor Model was not premature. Indeed it enjoyed initial popularity and underwent steady development.

However, Stent in his original article also referred to a development as premature such that when it occurred contemporaries did not adopt it by consensus. This is what happened with the Actor Model partly for the following reasons:

- For over 30 years after the first publication of the Actor Model, widely deployed computer architectures developed in the direction of making a single sequential thread of execution run faster.
- For over 25 years after the first publication, there was no agreed standard by which software could communicate high level data structures across organizational boundaries.

### Before its time?

According to [Gerson 2002], phenomena that lead people to talk about discoveries being before their time can be analyzed as follows:

> *We can see the phenomenon of 'before its time' as composed of two separate steps. The first takes place when a new discovery does not get tied to the conventional knowledge of its day and remains unconnected in the literature. The second step occurs when new events lead to the 'rediscovery' of the unconnected results in a changed context that enables or even facilitates its connection to the conventional knowledge of the rediscovering context.*

But circumstances have radically changed in the following ways:

- Progress on improving the speed of a single sequential thread has stalled for some time now. Increasing speed depends on effectively using many-core architectures.
- Better ways have been implemented that Actors can use to communicate messages between computers.
- Actors have been increasingly adopted by industry.

Consequently, by the criteria of Gerson, the Actor Model might be described by some as *before its time*.

According to [Zuckerman and Lederberg 1986], premature discoveries are those that were made but neglected. [Gerson 2002] argued,

> *But histories and sociological studies repeatedly show that we do not have a discovery until the scientific community accepts it as such and stops debating about it. Until then the proposed solution is in an intermediate state.*"

By his argument, the Actor Model is a discovery but since its practical importance is not yet accepted by consensus, its practical importance is not yet a discovery.

# Index

**End Notes**

is discrete. Surprisingly [Clinger 1981; later generalized in Hewitt 2006] answered the question in the negative by giving a counterexample:

Any finite set of events is consistent (the activation order and all reception orders are discrete) and represents a potentially physically realizable situation. But there is an infinite set of sentences that is inconsistent with the discreteness of the combined order and does not represent a physically realizable situation.

The resolution of the problem is to take discreteness of the combined order as an axiom of the Actor model:[4]

$$\forall[e_1, e_2 \in \text{Events}] \rightarrow \text{Finite}[\{e \in \text{Events} \mid e_1 \curvearrowright e \curvearrowright e_2\}]$$

Properties of concurrent computations can be proved using the above orderings [*e.g.* Bost, Mattern, and Tel 1995; Lamport 1978, 1979].

[5] The receiver might be on another computer and in any the system can make use of threads, locks, location transparency, throttling, load distribution, persistence, automated storage reclamation, queues, cores, channels, ports, *etc.* as it sees fit.

Messages in the Actor model are generalizations of packets in Internet computing in that they need not be received in the order sent. Not implementing the order of delivery, allows packet switching to buffer packets, use multiple paths to send packets, resend damaged packets, and to provide other optimizations.

For example, Actors are allowed to pipeline the processing of messages. What this means is that in the course of processing a message m1, an Actor can designate how to process the next message, and then in fact begin processing another message m2 before it has finished processing m1. Just because an Actor is allowed to pipeline the processing of messages does not mean that it *must* pipeline the processing. Whether a message is pipelined is an engineering tradeoff.

[6] The amount of effort expended depends on circumstances.

[7] These laws can be enforced by a proposed extension of the X86 architecture that will support the following operating environments:

- CLR and extensions (Microsoft)
- JVM (Oracle, IBM, SAP)
- Dalvik (Google)

Many-core architecture has made the above extension necessary in order to provide the following:

- concurrent nonstop automated storage reclamation (garbage collection) and relocation to improve performance,
- prevention of memory corruption that otherwise results from programming languages like C and C++ using thousands of threads in a process,
- nonstop migration of iOrgs (while they are in operation) within a computer and between distributed computers

[18] An example of the global state model is the Abstract State Machine (ASM) model [Blass, Gurevich, Rosenzweig, and Rossman 2007a, 2007b; Glausch and Reisig 2006].

[19] The lambda calculus can be viewed as the earliest message passing programming language [Hewitt, Bishop, and Steiger 1973] building on previous work.

For example, the lambda expression below implements a tree data structure when supplied with parameters for a leftSubTree and rightSubTree. When such a tree is given a parameter message "getLeft", it returns leftSubTree and likewise when given the message "getRight" it returns rightSubTree:

λ[leftSubTree, rightSubTree]
    λ[message]  message ◆ "getLeft" ⸲ leftSubTree
                          "getRight" ⸲ rightSubTree

[20] Allowing assignments to variables enabled sharing of the effects of updating shared data structures but did not provide for concurrency.

[21] There are nondeterministic computable functions on integers that cannot be implemented using the nondeterministic lambda calculus, i.e., using purely functional programming. By the Computational Representation Theorem, computations of effective Actor systems on integers are enumerable by a lambda expression. Consequently, every deterministic computable function on integers can be implemented by a lambda expression.

[22] [Petri 1962]

[23] [Close 2008]

[24] [Karp and Li 2007]

[25] which may require using membranes [Donnelley 1976, Hewitt 1980]

[26] *cf.* [Karp and Li 2008]

[27] [Hewitt, Bishop, and Steiger 1973, Hewitt and Baker 1977, Hewitt, Attardi, and Lieberman 1979]

[28] Consequently in Simula-76 there was no required locality of operations unlike the laws for locality in the Actor mode [Baker and Hewitt 1977].

[29] The ideas in Simula became widely known by the publication of [Dahl and Hoare 1972] at the same time that the Actor model was being invented to formalize concurrent computation using message passing [Hewitt, Bishop, and Steiger 1973].

[30] The development of Planner was inspired by the work of Karl Popper [1935, 1963], Frederic Fitch [1952], George Polya [1954], Allen Newell and Herbert Simon [1956], John McCarthy [1958, *et. al.* 1962], and Marvin Minsky [1968].

[31] This turned out later to have a surprising connection with Direct Logic. See the Two-Way Deduction Theorem below.

[32] Subsequent versions of the Smalltalk language largely followed the path of using the virtual methods of Simula in the message passing structure of

programs. However Smalltalk-72 made primitives such as integers, floating point numbers, etc. into objects. The authors of Simula had considered making such primitives into objects but refrained largely for efficiency reasons. Java at first used the expedient of having both primitive and object versions of integers, floating point numbers, etc. The C# programming language (and later versions of Java, starting with Java 1.5) adopted the more elegant solution of using boxing and unboxing, a variant of which had been used earlier in some Lisp implementations.

[33] According to the Smalltalk-72 Instruction Manual [Goldberg and Kay 1976]:

> There is not one global message to which all message "fetches" (use of the Smalltalk symbols eyeball, ◀; colon, ⁚; and open colon, ⁛) refer; rather, messages form a hierarchy which we explain in the following way-- suppose I just received a message; I read part of it and decide I should send my friend a message; I wait until my friend reads his message (the one I sent him, not the one I received); when he finishes reading his message, I return to reading my message. I can choose to let my friend read the rest of my message, but then I cannot get the message back to read it myself (note, however, that this can be done using the Smalltalk object *apply* which will be discussed later). I can also choose to include permission in my message to my friend to ask me to fetch some information from my message and to give that in information to him (accomplished by including ⁚ or ⁛ in the message to the friend). However, anything my friend fetches, I can no longer have.

In other words,

1) An object (let's call it the CALLER) can send a message to another object (the RECEIVER) by simply mentioning the RECEIVER's name followed by the message.

2) The action of message sending forms a stack of messages; the last message sent is put on the top.

3) Each attempt to receive information typically means looking at the message on the top of the stack.

4) The RECEIVER uses the eyeball, ◀ the colon, ⁚, and the open colon, ⁛, to receive information from the message at the top of the stack.

5) When the RECEIVER completes his actions, the message at the top of the stack is removed and the ability to send and receive messages returns to the CALLER. The RECEIVER may return a value to be used by the CALLER.

6) This sequence of sending and receiving messages, viewed here as a process of stacking messages, means that each message on the stack has a CALLER (message sender) and RECEIVER (message receiver). Each time the RECEIVER is finished, his message is removed from the stack and the CALLER becomes the current RECEIVER. The now current RECEIVER can continue reading any information remaining in his message.

7) Initially, the RECEIVER is the first object in the message typed by the programmer, who is the CALLER.

8) If the RECEIVER's message contains an eyeball, ◁; colon, ⁝, or open colon, ⁞, he can obtain further information from the CALLER's message. Any information successfully obtained by the RECEIVER is no longer available to the CALLER.

9) By calling on the object *apply,* the CALLER can give the RECEIVER the right to see all of the CALLER's remaining message. The CALLER can no longer get information that is read by the RECEIVER; he can, however, read anything that remains after the RECEIVER completes its actions.

10) There are two further special Smalltalk symbols useful in sending and receiving messages. One is the keyhole, ⫙, that lets the RECEIVER "peek" at the message. It is the same as the ⁞ except it does not remove the information from the message. The second symbol is the hash mark, **#**, placed in the message in order to send a reference to the next token rather than the token itself.

[34] The sender is an intrinsic component of communication in the following previous models of computation:

- *Petri Nets*: the input places of a transition are an intrinsic component of a computational step (transition).
- *Lambda Calculus*: the expression being reduced is an intrinsic component of a computational step (reduction).
- *Simula*: the stack of the caller is an intrinsic component of a computation step (method invocation).
- *Smalltalk 72*: the invoking token stream is an intrinsic component of a computation step (message send).

[35] An Actor can have information about other Actors that it has received in a message about what it was like when the message was sent. See section of this paper on unbounded nondeterminism in ActorScript.

[36] Arbiters render meaningless the states in the Abstract State Machine (ASM) model [Blass, Gurevich, Rosenzweig, and Rossman 2007a, 2007b; Glausch and Reisig 2006].

[37] The logic gates require suitable thresholds and other parameters.

It is possible for the following program to copy both of its input tapes onto its working tape:

*Step 1*: *Either*

    1.  copy the next input from the $1^{st}$ input tape onto the working tape and next do *Step 2,*

   *or*

    2.  copy the next input from the $2^{nd}$ input tape onto the working tape and next do *Step 3*.

*Step 2*: Next do *Step 1*.
*Step 3*: Next do *Step 1*.

It is also possible that the above program does not read any input from the $1^{st}$ input tape (*cf.* [Knabe 1993]). Bounded nondeterminism was but a symptom of deeper underlying issues with Nondeterministic Turing Machines.

[66] which can be optimized by reusing the thread if another message is waiting

[67] a globally unique identifier can be a 128-bit guid, long integer, or a string.

[68] Also, a reference for an Orleans Actor can be created from a C# *anObjectAddress* using
*aFactory*.CreateObjectReference(*anObjectAddress*).

[69] There can be optimizations for determinate message passing, *i.e.*, the same message always responds with the same result.

[70] Because of the ability to instantiate an Actor from its globally unique identifier, Orleans Actors are called "*virtual*" in their documentation. By analogy with virtual memory, the term "virtual" applied to an Orleans Actor would seem to imply that it would have to return to where it left. However, this terminology is misleading because an Actor can potentially migrate elsewhere and never come back.
     Better terminology would be to say that an Orleans Actor is "*perpetual.*"

[71] unless it is deleted by potentially unsafe means, which can result in a dangling globally unique identifier.

[72] after it has been unused for a while, its storage can be moved elsewhere outside the process in which it currently resides

[73] unless it is deleted by potentially unsafe means, which can result in a dangling globally unique identifier.

[74] However, after the message is finished processing, sometimes waiting method calls it invoked can be processed concurrently if they are independent.

[75] provided that the Actor is not contended

[76] [Microsoft 2013]

[77] reentrancy allows execution of waiting method calls to be freely interleaved

[78] [Liskov and Shira 1988; Miller, Tribble, and Shapiro 2005]

[79] Orleans uses Task<*aType*> for the type of a promise which corresponds to the type Future◁*aType*▷ in ActorScript.

[80] for requests, *e.g.*, method calls. Customers are sometimes called continuations in the literature although continuations often cannot handle exceptions.

[81] However, Orleans does still surfaces customers using lower level primitives.

[82] [ECMA 2014]

[83] [Barton 2014]

[84] analogous the await primitive in C# [Microsoft 2013]